\documentclass[aps, pra, superscriptaddress, letterpaper,twocolumn,showpacs,nofootinbib]{revtex4-1}
\usepackage{amsbsy,amssymb,amsmath,bm,bbold}
\usepackage{mathptmx,newtxtext,newtxmath,xspace}
\usepackage{graphicx,color,epsfig,rotate}
\usepackage{fancyhdr}
\usepackage{gensymb}
\usepackage[colorlinks=true, 
                    linkcolor=blue, 
                    urlcolor=blue,
                    citecolor=blue]{hyperref}

\usepackage[normalem]{ulem}
\newcommand{\stkout}[1]{\ifmmode\text{\sout{\ensuremath{#1}}}\else\sout{#1}\fi}

\def\bbbc{{\mathchoice {\setbox0=\hbox{$\displaystyle\rm C$}\hbox{\hbox
to0pt{\kern0.4\wd0\vrule height0.9\ht0\hss}\box0}}
{\setbox0=\hbox{$\textstyle\rm C$}\hbox{\hbox
to0pt{\kern0.4\wd0\vrule height0.9\ht0\hss}\box0}}
{\setbox0=\hbox{$\scriptstyle\rm C$}\hbox{\hbox
to0pt{\kern0.4\wd0\vrule height0.9\ht0\hss}\box0}}
{\setbox0=\hbox{$\scriptscriptstyle\rm C$}\hbox{\hbox
to0pt{\kern0.4\wd0\vrule height0.9\ht0\hss}\box0}}}}

\newcommand{\beq}{\begin{equation}}
\newcommand{\eeq}{\end{equation}}
\newcommand\bea{\begin{eqnarray}}
\newcommand\eea{\end{eqnarray}}
\newcommand\ba{\begin{array}}
\newcommand\ea{\end{array}}
\newcommand{\nn}{\nonumber}

\definecolor{purple}{cmyk}{ 0.5, 0.7, 0,0}
\definecolor{magenta}{cmyk}{ 0, 1, 0,0}

\def\be{\begin{equation}}
\def\ee{\end{equation}}
\def\bea{\begin{eqnarray}}
\def\eea{\end{eqnarray}}

 \allowdisplaybreaks[2]

\begin{document}
\title{Chiral liquid phase of simple quantum magnets}
\author{Zhentao~Wang}
\affiliation{Department of Physics and Astronomy, The University of Tennessee, Knoxville, Tennessee 37996, USA}
\author{Adrian~E.~Feiguin}
\affiliation{Department of Physics, Northeastern University, Boston, Massachusetts 02115, USA}
\author{Wei~Zhu}
\affiliation{T-4 and CNLS, Los Alamos National Laboratory, Los Alamos, NM 87545, USA}
\author{Oleg~A.~Starykh}
\affiliation{Department of Physics and Astronomy, University of Utah, Salt Lake City, Utah 84112, USA}
\author{Andrey~V.~Chubukov}
\affiliation{Department of Physics and William I. Fine Theoretical Physics Institute, University of Minnesota, Minneapolis, Minnesota 55455, USA}
\author{Cristian~D.~Batista}
\affiliation{Department of Physics and Astronomy, The University of Tennessee, Knoxville, Tennessee 37996, USA}
\affiliation{Quantum Condensed Matter Division and Shull-Wollan Center, Oak Ridge National Laboratory, Oak Ridge, Tennessee 37831, USA}
\date{\today}

\begin{abstract}
We study a $T=0$ quantum phase transition between a quantum paramagnetic state and a magnetically ordered state for a spin
$S=1$ XXZ Heisenberg antiferromagnet  on a two-dimensional triangular lattice. The transition is induced by an
easy plane single-ion anisotropy $D$. At the mean-field level, the system undergoes a direct transition at a critical
$D = D_c$ between a paramagnetic state at
$D > D_c$ and an ordered state with broken U(1) symmetry at $D < D_c$.
We show that beyond mean field the phase diagram is very different and includes an intermediate, partially ordered {\em chiral liquid} phase.
Specifically, we find that
inside the paramagnetic phase the Ising ($J_z$) component of the Heisenberg exchange
 binds magnons into a two-particle bound state
 with zero total momentum and spin.
This bound state condenses
  at $D > D_c$, {\em before} single-particle excitations become unstable,
and gives rise to a chiral liquid phase, which spontaneously breaks spatial inversion symmetry, but leaves the spin-rotational U(1)
 and time-reversal symmetries intact. This chiral liquid phase is characterized by a finite vector chirality
without long range dipolar magnetic order.
 In our analytical treatment, the chiral phase appears for arbitrarily small  $J_z$  because the magnon-magnon attraction becomes singular
 near the single-magnon condensation transition.
This phase exists in a finite range of  $D$ and   transforms into the magnetically ordered state
at some $D<D_c$. We corroborate our analytic treatment with numerical
 density matrix renormalization group calculations.
 \end{abstract}

\pacs{ }

\maketitle %
\thispagestyle{fancy}

\section{Introduction}

Broken symmetries are ubiquitous in nature.
 Many broken-symmetry states have conventional long-range orders, such as dipolar magnetism or charge/orbital order, but some have more complex
 {\em composite} orders with order parameters built out of {\em non-linear}
combinations of the original spin degrees of freedom.  An example of such order is a spin nematic,
 whose order parameter
 is a bilinear combination of spin operators ~\cite{Andreev1984,
  chubukov1990,Penc2011}. Bilinear
 order parameters often emerge in frustrated spin
systems, such as
  $J_1$-$J_2$-$J_3$ Heisenberg model on a square lattice \cite{Chandra1990,Chandra1990a,Capriotti2004}, and describe spontaneous breaking of a discrete lattice rotational symmetry
while spin-rotational  SU(2) symmetry remains unbroken.

One of the first studies of composite orders was performed by Villain \cite{Villain1977},  who
considered helical (spiral) spin order in Heisenberg and XY spin models in
 an external magnetic field ${\bm h}= h {\hat {\boldsymbol  z}}$. He noticed that a helical order
breaks both  continuous and discrete symmetries.  The continuous symmetry breaking corresponds
to the development of a conventional dipolar magnetic order in the direction perpendicular to the field, i.e.,
to a finite expectation value $\langle {\bm{S}^{x,y}_{\bm{n}}}\rangle$ of the spin operator at every site of the lattice. The
 discrete symmetry breaking distinguishes between clockwise and anticlockwise rotations of spins from site ${\bm n}$
to site ${\bm m}$ along the bond $\langle {\bm n},{\bm m}\rangle$. Such an order is chiral in nature and  the corresponding order parameter, {\em vector chirality},  is the $z$ component of the vector product
of spins on a given bond ${\kappa}_{\bm{nm}} =  {\hat {\boldsymbol  z}} \cdot {\bm{S_n}} \times \bm{S_m}$.

The fact that both continuous and discrete symmetries are broken in the ordered phase ($\langle \bm{S_n}\rangle \neq 0$ and $\langle \kappa_{\bm{nm}} \rangle \neq 0$) opens up a possibility of a {\em sequence of phase transitions}
 between this phase and the paramagnetic one, where $\langle \bm{S_n}\rangle = 0$ and $\langle \kappa_{\bm{nm}}  \rangle= 0$.
 In the context of classical helimagnetism, Villain argued~\cite{Villain1977,Villain1978} that
 $\langle \bm{S_n}\rangle$ and $\langle \kappa_{\bm{nm}} \rangle$  do not need to acquire
finite values simultaneously and
 that the paramagnetic and the magnetically ordered phases
 may be separated by the novel {\em chiral liquid} (CL) phase, in which
 the  chiral order parameter is finite, i.e.  $ \langle \kappa_{\bm{nm}} \rangle \neq 0$, but long-range magnetic order is absent
  ($\langle \bm{S_n}\rangle = 0$).
   A similar set of ideas has been recently applied to itinerant electron systems featuring various nematic orders \cite{Fradkin2010,Fernandes2012}.

For thermodynamic phase transitions
  in  U(1)-symmetric systems,
the CL is expected to exist in a finite-temperature window $T_{\rm mag} < T < T_{\rm ch}$,
where $T_{\rm ch}$ is
the onset  temperature of
long-range  chiral
 order and $T_{\rm mag}$ is the onset temperature
  of  long-range magnetic order
(Berezinskii-Kosterlitz-Thouless quasi-long-range order in two dimensions).
Numerous numerical studies of two-dimensional classical helimagnets
~\cite{Kawamura1998,Hasenbusch2005,Korshunov2006,Sorokin2012,Schenck2014} have  found that
$ T_{\rm ch}$ and $T_{\rm mag}$ are indeed different, but
the relative difference
is very small,
at best only a few percent.

 Here, we consider a {\em quantum phase transition}
  at $T=0$
    in
    systems with U(1) spin symmetry,
     driven by quantum fluctuations \cite{Sachdev2001}.  Our goal is to understand whether a CL state can emerge as the ground state of the quantum spin system, separating a  quantum paramagnet
from a magnetically ordered phase.
We argue below that the minimal model that describes this physics is a spin-1 triangular lattice XXZ antiferromagnet with nearest-neighbor exchange $J$ and single-ion anisotropy $D$.
 We find that CL phase exists in a rather wide range of $D$, whose width can be as large as $J/2$.

\begin{figure}[!tbp]
	\centering
	\includegraphics[width=\columnwidth]{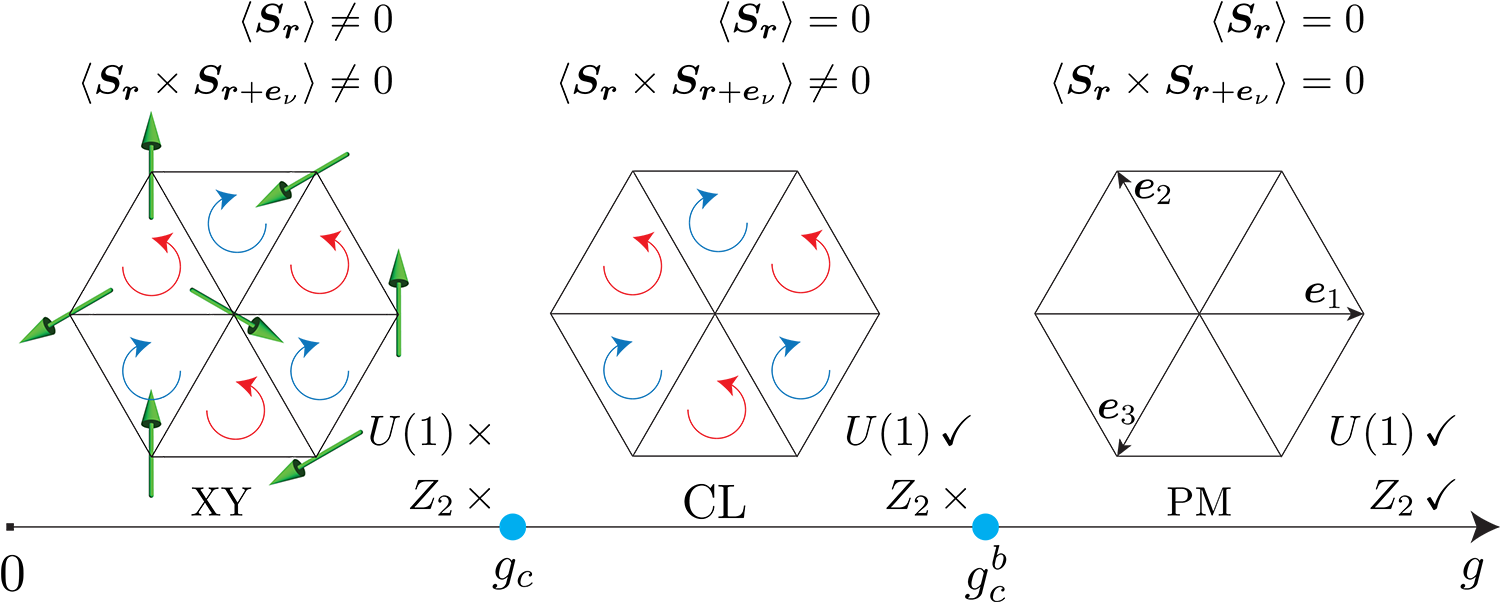}
	\caption{A schematic plot of the three phases \{XY, CL, PM\} as a function of the tuning parameter $g$, where $g \sim D/J$ for the quantum spin-1
		model considered in this paper. For similar sequence of transitions in classical models $g \sim T$.
		The symmetries of each phase are noted on the bottom right of each hexagon.
	}
	\label{fig:schematic}
\end{figure}

Our key result is presented in Fig.~\ref{fig:schematic}. It shows that a featureless paramagnetic state, realized at $D \gg J$, is separated from the magnetically ordered
XY state at $D \ll J$ by the intermediate CL phase, which is stable in the finite window $D_c < D < D_c^b$.

In more specific terms, we analyzed the effects of magnon-magnon interaction in the paramagnetic phase.  There are two gapped magnon modes in this phase.
Their dispersion has minima at $\pm {\bm Q}$, where ${\bm Q} = (4\pi/3,0)$.
Within self-consistent mean-field theory, magnon excitations soften at these momenta at $D_c \approx 2.68J$, and at smaller $D$ the system has an XY long-range spiral magnetic order, which breaks  continuous U(1)
symmetry (long-range magnetic order) and  discrete $Z_2$
  (chiral symmetry, spatial inversion, or parity).
     We found that the interaction between magnons  with opposite spins is attractive.
       This attraction leads to the formation of a pair condensate of two magnons with zero
        total spin  and zero total momentum, while individual magnon momenta
       are near $\pm {\bm Q}$.   The attraction comes from the Ising, $J_z$, part of the exchange interaction, and involves both
``normal'' interaction terms with two creation and two annihilation magnon operators and ``anomalous" interaction terms which do not conserve magnon numbers.
The
  pairs condense  at $D = D^b_c > D_c$, when single-magnon excitations are still gapped.  The
   condensation  gives rise to a finite
 staggered
 vector chirality $\kappa \neq 0$ for each elementary triangle of spins (up- and down-pointing triangles have opposite chirality).  This CL state spontaneously breaks spatial inversion (parity) symmetry  but preserves the time-reversal and translational symmetries
 because each unit cell includes one up-pointing and one down-pointing triangle.

The width of the chiral phase depends on the value of $J_z$. 
In our perturbative treatment, we  found that there is no threshold of $J_z$, i.e.,  CL state develops  for arbitrary small $J_z$ because the pairing interaction is singular at $ D = D_c$.
This singularity appears at second order in $J_z$ due to strong quantum renormalization of the interaction between magnons.
Known from the previous renormalization group analysis, this spontaneous breaking of continuous U(1) and discrete $Z_2$ symmetries would become weakly first order, leading to a small but finite critical $J_z$~\cite{Kamiya2011,Parker2017}. 

The paper is organized as follows. In  Sec.~\ref{sec:model_1} we introduce the model
(Sec.~\ref{sec:model})
 and consider a toy problem of a
 single  two-spin bound state (Sec.~\ref{sec:exciton}).
  In
Sec.~\ref{sec:bosons} we introduce  Schwinger bosons and solve for
the two-magnon  bound state in a many-body system in the presence of  quantum fluctuations.
 We
 first present self-consistent analysis of single-magnon excitations and find the critical  $D_c$.
  (Sec.~\ref{sec:bosons_1}).
  Then, we derive the interaction between low-energy magnons
   (Sec.~\ref{sec:interaction}),
    and show (Sec.~\ref{sec:eom}) that  it is attractive in the channel, where the condensation of a two-magnon bound state leads to vector chirality.  We solve for the two-magnon bound state first at small $J_z$ (Sec.~\ref{sec:eom_1}), to order  $J^2_z$, and then for a generic $J_z$
   (Sec.~\ref{sec:BS}).
 In Sec.~\ref{sec:dmrg} we present DMRG calculations, which support our analytical results.
 In Sec.~\ref{sec:summary}  we
  summarize
  our findings
   and outline
    their connections with other physical systems
of current interest.
 In Appendix~\ref{sec:Ising} we discuss two other phases in the vicinity of the CL phase: an Ising spin-density wave state and a supersolid (SS) state.
 In Appendix~\ref{sec:hard-core-boson}
 we present the analysis of the bound-state development using an alternative procedure to relate spin operators to bosons.

{\it Relation to earlier works. }
The separation between the breaking of a continuous and a discrete symmetry,  either in classical (thermodynamic) or in quantum phase transitions, has been discussed
 for various physical problems.  Several Heisenberg spin models on a square lattice,
   e.g., $J_1$-$J_2$ model, Ref. \cite{Chandra1990,Chandra1990a}, and $J_1$-$J_3$ model, Ref.  \cite{Capriotti2004}, display  $T=0$ order
   which breaks not only the SU(2) spin-rotational symmetry, but also a discrete lattice rotational symmetry. Thus, the ground state of the $J_1$-$J_2$ model at large $J_2$ is a stripe order with ferromagnetic spin arrangement either along $X$ or along $Y$ spatial direction.
The
 order parameter associated  with the difference between $X$ and $Y$ directions (an Ising nematic order)
is {\em quadratic} in the spin operators
~\cite{Chandra1990}.
In two dimensions (2D), spin-rotational symmetry
 cannot be broken at any finite temperature $T\neq 0$, so that $\langle {\bm S}\rangle = 0$,
but the discrete $C_{\rm 4}$ lattice rotational symmetry  breaks spontaneously  down to $C_{\rm 2}$ below a certain Ising transition temperature $T_{\rm{Ising}}$.
This leads to a finite-temperature
 liquid
  nematic phase with  broken  $Z_2$ symmetry. This
has been
 identified in numerical
  studies~\cite{Weber2003,Capriotti2004}.

In three dimensions (3D), long-range magnetic order $\langle {\bm S}\rangle \neq 0$ is present below N\'eel temperature $T < T_N$, but still there exists a temperature interval $T_N < T < T_{\rm{Ising}}$
  where only a
  nematic order is present.
   For itinerant fermion systems, these ideas formed the basis~\cite{Fernandes2012}
    for the magnetic scenario of the nematic order, observed in Fe-based superconductors.

In one-dimensional(1D) systems, continuous symmetries are preserved even at $T=0$ because of  the singular nature of quantum fluctuations~\cite{Sachdev2001}. There have been several studies of composite vector chiral (VC) orders at $T=0$.
A spin chiral order with orbiting spin currents was found in $S=1/2$ two-leg zigzag Heisenberg spin ladder with XXZ-type exchange interaction~\cite{Nersesyan1998}.
For an isotropic Heisenberg spin chain with competing interactions, it was shown \cite{Kolezhuk2005} that an
 external magnetic  field acts in the same way as an exchange anisotropy  and stabilizes long-ranged chiral order \cite{chubukov1991_1,Hikihara2008}.
 A chiral order has been also found in a two-leg fully frustrated Bose-Hubbard ladder \cite{Dhar2013} and was argued to generate staggered orbital currents circling around elementary plaquettes~\cite{Nersesyan1991,Fjaerestad2002}. Chiral phases have also been observed in $S=1$ zig-zag ladder \cite{Hikihara2001,Greschner2013}.
Magnetically-ordered states
coexisting with
 chiral orders have also been studied at $T=0$ in triangular \cite{Momoi1997,Chubukov2013} and kagom\'e \cite{Domenge2005} geometries.

As described above, a VC order  ${\kappa}_{\bm{nm}} \neq 0$ spontaneously  breaks parity (a symmetry with respect to  spatial inversion),
but preserves time-reversal symmetry. This makes VC order, which is the topic of our study, very different from
 {\em scalar chiral} order
 $\chi_{\bm{nml}} = \bm{S_n} \cdot \bm{S_m} \times \bm{S_l}$ (where sites ${\bm n}, {\bm m}$ and ${\bm l}$ form, e.g., a triangular plaquette).
Such an order breaks both  parity and time-reversal symmetries \cite{Wen1989}.
A ground state with a scalar chiral order without usual long-ranged magnetic order was proposed at the beginning of high-$T_{\rm c}$ era by Kalmeyer and Laughlin \cite{Kalmeyer1987,Kalmeyer1989},
who used a quantum-Hall-like incompressible bosonic wave function
to describe it. In close analogy with the quantum Hall effect, this chiral spin state has  gapped excitations in the bulk but gapless  excitations at the edge of a sample.
After almost 20 years this proposal has received a confirmation in a series of recent analytical and numerical studies of $S=\frac{1}{2}$ 
antiferromagnets~\cite{Schroeter2007,Yao2007_chiral_spin_liquid,Ronny2009,Gong2014,Zhu2014_kagome,He2014_kagome,He2015_kagome,Zhu2015_kagome}.


Chiral (noncentrosymmetric) itinerant helimagnets have been found to exhibit a first-order thermodynamic phase transition into a chiral liquid phase that preempts the onset of magnetic ordering~\cite{Tewari2006}. While this phase does not break the chiral symmetry spontaneously, it shows that the chiral susceptibility can diverge while the magnetic susceptibility is still finite.

Our finding of the two-dimensional CL phase with finite vector chirality and no dipolar magnetic order is a realization of the composite VC order in the ground state of a two-dimensional quantum spin model.

\section{Anisotropic $S=1$ Triangular Antiferromagnet}
\label{sec:model_1}
\subsection{Spin-1 model}
\label{sec:model}
We consider  $S=1$
 model on a triangular lattice, with anisotropic XXZ antiferromagnetic exchange  between nearest neighbors and an
easy-plane single-ion anisotropy, $D (S^z)^2$,  with $D>0$. This is the minimal model  to study a quantum phase transition between a  quantum paramagnet and a magnetically ordered state with
 an
  additional discrete symmetry breaking.
   Despite simplicity, the model  describes real materials~\cite{Zapf06,Zapf14}.

The
Hamiltonian
of the model
is
\begin{equation}
\label{eq:ham}
{\cal H} = \sum_{{\bm r},\nu, \mu} J_{\mu} S^{\mu}_{\bm r} S^{\mu}_{{\bm r}+{\bm e}_{\nu}}
+D \sum_{{\bm r}} (S^z_{{\bm r}})^2,
\end{equation}
where
${\bm e}_1= a {\hat {\bm x}}$, ${\bm e}_2=a (-{\hat {\bm x}}/2 + \sqrt{3}{\hat {\bm y}}/2) $ and ${\bm e}_3= a (-{\hat {\bm x}}/2 - \sqrt{3}{\hat{\bm  y}}/2)$ (see Fig.~\ref{fig:schematic}),
$a$ is the lattice constant, $\mu=\{x, y, z\}$, $J_x=J_y = J$ and $J_z= \zeta J$.
We keep $\zeta$ of order one through most of the paper, but
 will
 consider
 the
 limits of small $\zeta$ in Sec.\ref{sec:eom} and large $\zeta$ in Appendix. \ref{sec:Ising}.

 The
   model of Eq. \eqref{eq:ham}
    has  two distinct phases at small and at large $D$.  At $D=0$, it reduces to a U(1)-symmetric XXZ Heisenberg model on a triangular lattice, which develops
    a $120\degree$ three-sublattice
 long-range magnetic  order at $T=0$.
   Aside from breaking
the continuous U(1) symmetry of global spin rotations along the $z$ axis, this non-collinear ordering also breaks the discrete
chiral symmetry. The sign of $\langle \kappa_{\bm{nm}} \rangle$ is positive
 for the ground state in which
  the angle between
the spins at sites ${\bm n}$ and ${\bm m}$ is $2\pi/3$ and negative for the alternative ground state in which the angle is $-2\pi/3$.

In contrast, the ground state for large enough $D$ is a
 magnetically disordered state in which
 each spin is in the $|S^z=0 \rangle$
  configuration with
   $\langle {\bm S} \rangle =0$,   $\langle  S^2_z \rangle =0$, $\langle  S^2_x \rangle  = \langle  S^2_y \rangle =1$.
This product-like
state preserves time-reversal and all lattice symmetries of the model, and therefore represents
a featureless quantum paramagnet.

The goal of our work is to understand whether an intermediate chiral liquid (CL) phase exists between a quantum paramagnet and
a  magnetically ordered state at $T=0$.

\subsection{Toy problem of a two-spin bound state}
\label{sec:exciton}

To develop physical intuition, we first consider a toy problem of 
the bound-state formation for two magnons excited above the ground state at large $D$. The magnons carry opposite spins $S^z = \pm 1$,
so that the total spin of such a two-spin ``exciton" is zero. Its wave function is written as
\begin{equation}
\label{eq:o1}
|{\rm ex}\rangle \!= \!\! \sum_{n\neq m} \psi_{n,m} |n,m\rangle ~\text{where}~ |n,m\rangle \!=\! \frac{1}{2} S^+_n S^{-}_m ~\otimes_j |0\rangle_j .
\end{equation}
Here, $|0\rangle_j$ denotes the $|S^z_j=0 \rangle$ state at site $j$.
Projecting  ${\cal H} |{\rm ex}\rangle = E |{\rm ex}\rangle$ onto a single exciton subspace,
 we obtain
  an effective Schr\"{o}dinger equation for the pair wave function $\psi_{n,m}$:
\begin{align}
&\quad (E-2D) \psi_{n,m} \nonumber \\
& = J \sum_g \left[ \psi_{n+g,m} + \psi_{n,m+g} - \zeta \psi_{n,m} \delta_{n,m+g} \right],\label{eq:o2}
\end{align}
where
$g=\{ \bm{e}_1, \bm{e}_2, \bm{e}_3, -\bm{e}_1, -\bm{e}_2, -\bm{e}_3 \}$
 runs over six
 nearest neighbors.

Evidently, the last term of this equation describes an attraction between the state with $|S^z_n =+1 \rangle$ at site $n$ (a particle) and the state with $|S^z_m=-1 \rangle$
  at a neighboring site $m = n-g$ (a hole).  Fourier transforming into momentum space, we obtain that the wave function of a ``particle-hole'' pair with the center-of-mass (c.m.)
momentum ${\bm K}$ and the relative momentum ${\bm q}$:
\begin{equation}
\label{eq:o3}
\Psi_{\bm K}({\bm q}) = \frac{1}{N}\sum_{n,m} e^{i {\bm K}\cdot ({\bm n} + {\bm m})/2} e^{i {\bm q}\cdot ({\bm n} - {\bm m})} \psi_{n,m}
\end{equation}
obeys the following integral equation:
\begin{align}
\label{eq:o4}
&\quad \! \left[ E \! - \! 2D \! - \! J \sum_g( e^{- i{\bm g}\cdot (\frac{1}{2} {\bm K} + {\bm q})} \! +\! e^{-i {\bm g}\cdot (\frac{1}{2} {\bm K} - {\bm q})}) \right] \Psi_{\bm K}({\bm q}) \nonumber\\
& = - \zeta J \sum_g e^{i {\bm g} \cdot {\bm q}} \frac{1}{N} \sum_{\bm p} e^{-i {\bm g} \cdot {\bm p}} \Psi_{\bm K}({\bm p}) \\
& \equiv - \zeta J \sum_g e^{i {\bm g} \cdot {\bm q}} B_{\bm g} .\nonumber
\end{align}
In the last line we introduced $B_{\bm g}$ via
\begin{equation}
\label{eq:o4_1}
 B_{\bm g} = \frac{1}{N} \sum_{\bm p} e^{-i {\bm g} \cdot {\bm p}} \Psi_{\bm K}({\bm p}).
\end{equation}
The left-hand side of  Eq. (\ref{eq:o4})  can be expressed as $ [E - \omega_{{\bm p}_1} - \omega_{{\bm p}_2}]  \Psi_{\bm{K}}(\bm{q})$, where
$\omega_{\bm p} = D + J \sum_{\bm g} e^{i {\bm p} \cdot {\bm g}} = D + 2J(\cos[p_x] + 2\cos[\frac{p_x}{2}] \cos[\frac{\sqrt{3} p_y}{2}])$
is the single particle dispersion. In these notations, a particle and a hole
carry
momenta ${\bm p}_1 = {\bm K}/2 + {\bm q}$
and ${\bm p}_2 = {\bm K}/2 - {\bm q}$. The right-hand side represents the Ising interaction between a particle and a hole, sharing the same bond of the lattice.

By standard manipulations, this equation is reduced to the matrix one
\begin{equation}
\label{eq:o5}
B_{\bm g} = \zeta J \sum_{{\bm g}'} M_{{\bm g} {\bm g}'} B_{{\bm g}'},
\end{equation}
where
  the kernel is
\begin{equation}
\label{eq:o6}
M_{{\bm g} {\bm g}'} = \frac{1}{N} \sum_{\bm q} \frac{e^{i {\bm q} \cdot ({\bm g}'-{\bm g})}}{4J \sum_{j=1}^3 \cos[\frac{K_j}{2}] \cos[q_j] + 2 D - E},
\end{equation}
and $q_j \equiv {\bm q} \cdot {\bm g}_j$.
  Solving Eq. ~\eqref{eq:o5}, we obtain the energy $E$
  of an exciton with the c.m. momentum ${\bm K}$.

We analyzed at what value of $D$ the exciton energy $E$ vanishes for various c.m. momenta ${\bm K}$  for a given $J_z = \zeta J$ and found the largest $D$ for  ${\bm K} = 0$.
For  this ${\bm K}$
 the minimum of $E
({\bm K}=0)$ occurs at ${\bm q} = \pm {\bm Q}$, and at the minimum $E^{\rm min}({\bm K}=0) = 2D - 6J$.
 We parametrize relative momenta as
 ${\bm q} = \pm {\bm Q} + {\bm p}$ and expand the denominator of Eq.~\eqref{eq:o6} in small  ${\bm p}$. This leads to
 \begin{equation}
 \label{eq:o7}
M_{{\bm g} {\bm g}'} \approx 2 \cos[{\bm Q}\cdot ({\bm g}'-{\bm g})] I_{0},
\end{equation}
where, to a logarithmic accuracy,
\begin{equation}
\label{eq:o8}
I_{0} = \frac{\sqrt{3}}{8\pi^2}\int d{\bm p} \frac{1}{\frac{3J}{2} (p_x^2 + p_y^2) + \epsilon_b} = \frac{\ln\Big(\frac{3J \Lambda^2}{2\epsilon_b}\Big)}{4\sqrt{3} \pi J}.
\end{equation}
Here, $\Lambda$ is the upper-momentum cutoff in the $p$ integration
and  $\epsilon_b = 2(D-3J) - E$  is the  binding energy of an exciton.
  With these simplifications,  Eq.~\eqref{eq:o5} turns into $\sum_{\bm g} ( \alpha \cos[{\bm Q}\cdot ({\bm g}'-{\bm g})] - \delta_{{\bm g} {\bm g}'}) B_{\bm g} = 0$,
where $\alpha = 2\zeta J I_{0}$, and, we remind, ${\bm g}=\{ \bm{e}_1, \bm{e}_2, \bm{e}_3, -\bm{e}_1, -\bm{e}_2, -\bm{e}_3 \}$ has six components, by the number of nearest neighbors.  This equation can be easily solved. The condition that the determinant
    vanishes
    yields the quadratic equation on $\alpha$: $ 1 - 6 \alpha + 27 \alpha^2/4 =0$. This equation  has two solutions: $\alpha_1 = \frac{2}{9}$ and $\alpha_2 = \frac{2}{3}$. The corresponding binding energies are $\epsilon_{b,\nu} = \frac{3}{2}J \Lambda^2 \exp[-2 \sqrt{3} \pi \alpha_\nu /\zeta ]$
    ($\nu =1,2$).
     Both are non-zero already at arbitrary small
 $\zeta = J_z/J$.
The exciton energy $E = 2D - 6J - \epsilon_{b, \nu}$ {\em vanishes} at a critical
$D= D_\nu$, where
\begin{equation}
\label{eq:o9}
D_\nu = 3J \left(1 + \frac{\Lambda^2}{4} e^{-\frac{2 \sqrt{3} \pi \alpha_\nu}{\zeta}}\right)
\end{equation}
We note that for both solutions $\alpha_{1}$ and $\alpha_2$  this happens when the minimum of  the
particle-hole
 continuum is still at a finite energy
 ($D-3J >0$).

\begin{figure}[!tbp]
  \centering
  \includegraphics[width=0.8\columnwidth]{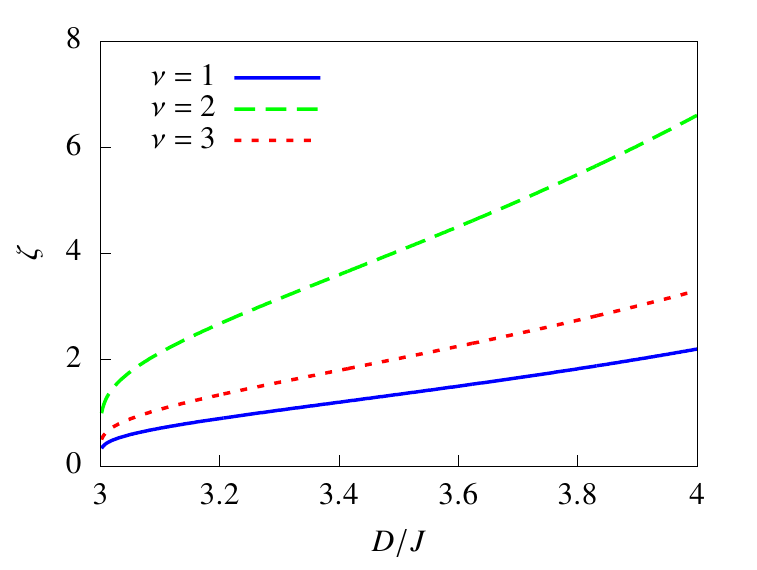}
  \caption{
 Plot of Eq.~\eqref{eq:o9} with three solutions $\alpha_1=\frac{2}{9}, \, \alpha_2=\frac{2}{3}$ (
 both
 for ${\bm K} = 0$), and $\alpha_3=\frac{1}{3}$ (for $\bm{K} = 2\bm{Q}$). The cutoff
 scale is set at $\Lambda=2$.
 }
  \label{fig:delta}
\end{figure}

Comparing the two solutions, we find that if we keep $J_z = \zeta J$ fixed and progressively reduce $D$ towards $3J$, the first instability occurs for the
 solution with $\alpha_{1} = \frac{2}{9}$.
    One can easily verify that the  eigenfunction
    $B_{\bm g}^{(1)}$
    for this solution is {\em odd} under spatial inversion ${\bm g}_j \to -{\bm g}_j$, i.e., viewed as a
  function of six elements of $g$, it behaves as
\begin{equation}
\label{eq:o10}
B_{\bm g}^{(1)} = (1,1,1,-1,-1,-1)^{\rm T}.
\end{equation}
This in turn implies that
that $\Psi_{{\bm K}=0}^{(1)}({\bm q})$ is an odd function of ${\bm q}$.
Solving the actual equation on $B_{\bm g}^{(1)}$
by expanding $E({\bm K} =0)$ about its minimum
and transforming
 from $B_{\bm g}$ to  $\Psi_{\bm K}({\bm p})$
  using the inverse of  Eq. (\ref{eq:o4_1}), we obtain
\begin{equation}
\label{eq:o11}
\Psi_{{\bm K}=0}^{(1)}({\bm q}) = \frac{-2i \zeta J(\sin[q_x]-2\sin[\frac{q_x}{2}] \cos [\frac{\sqrt{3}q_y}{2}])}{4J (\cos[q_x]+2\cos[\frac{q_x}{2}] \cos[\frac{\sqrt{3}q_y}{2}]) + 6J + \epsilon_{b,1}} .
\end{equation}
In real space,
   the corresponding $\psi^{(1)}_{n,m} = \psi({\bm r}={\bm n}-{\bm m}) \sim \sin[{\bm Q} \cdot {\bm r}] e^{-r/\xi}$. The size of the exciton scales as $\xi \sim \epsilon_{b,1}^{-1/2}$.

For other c.m. momenta, a two-spin exciton also develops, but its energy becomes negative at
a smaller $D < D_1$.
Thus, for ${\bm K} = 2{\bm Q} = 2 (4\pi/3,0)$, the minimum of the denominator in \eqref{eq:o7} occurs at ${\bm q}=0$.
For small $\bm{q}$, the eigenvalue equation yields a single root
$\alpha_3 = \frac{1}{3} > \alpha_1$.  This leads to the pair condensation at a smaller $D$  than for the parity-breaking
solution $\alpha = \alpha_1$ (see Fig.~\ref{fig:delta}).

 As a hint what $E <0$ means, consider a finite density of bound pairs. Once the
 pairs condense,
 the new ground state at $D < D_1$ can be described at a mean-field level by the Jastrow wave function~\cite{Zaletel2014}:
\begin{equation}
| \Psi_{\rm CL} \rangle \sim e^{u \sum_{\bm{k}} \phi(\bm{k}) S_{\bm{k}}^+ S_{\bm{k}}^-} | 0 \rangle,
\label{eq:CL}
\end{equation}
where the real function $\phi(\bm{k})=-\phi(-\bm{k})$ is odd under inversion, and $u$ is a real number. As a result,
$\Psi_{\rm CL}$ breaks
spatial inversion but preserves time-reversal symmetry. One can straightforwardly check that in this state
the $z$ component of vector chirality ${\kappa}_{{\bm n} {\bm m}}$ is finite on every bond $\langle {\bm n}, {\bm m}\rangle$ while
 $\langle {\bm S}_{\bm n} \rangle =0$ for every site ${\bm n}$.
Therefore, \eqref{eq:CL} describes the
 CL state.
The sign of $u$ selects the direction of the chiral order on a given bond and encodes $Z_2$ character of the CL phase.
This state is also called a {\em spin-current} state \cite{Chubukov2013} because bond variables
${\kappa}_{{\bm n} {\bm m}}$ form oriented closed loops on every elementary triangle of the lattice.

\section{Schwinger boson formulation}
\label{sec:bosons}

\subsection{SU(3) Schwinger boson representation}
\label{sec:bosons_1}
Having demonstrated the possibility of a VC order
 by analyzing the energy of a single two-spin exciton on top of the product $\otimes_j |S^z_j=0 \rangle$ state,
we now turn to the technical task of establishing its existence in the full many-body problem.
 For this,
  we need the formalism capable of treating both the large-$D$ paramagnetic
   state and the low-$D$
   magnetically ordered state. Such a formulation is provided by the Schwinger boson theory \cite{Auerbach1994} associated with the fundamental
representation of SU(3)~\cite{Muniz2014}. The bosons obey the constraint, which needs to be fulfilled at every site ${\bm r}$ of the lattice,
\begin{equation}\label{constraint}
\sum_m b^{\dagger}_{{\bm r}m} b^{\;}_{{\bm r}m}=1,
\end{equation}
with $m=\{\uparrow, 0, \downarrow \}$
 label
 eigenvectors of $S^z_{\bm r}$:
$S^z_{\bm r} b^{\dagger}_{{\bm r}m}|0\rangle = s^z_m b^{\dagger}_{{\bm
r}m}|0\rangle$ with $s^z_{\uparrow}=1,s^z_{0}=0$, and $s^z_{\downarrow}=-1$.
 We will enforce the constraint in Eq.~\eqref{constraint} by introducing the Lagrange multipliers $\mu_{\bm r}$:
\begin{eqnarray}
{\bar {\cal H}} = {\cal H} + \sum_{\bm r} \mu_{\bm r}
(b^{\dagger}_{{\bm r}\uparrow} b_{{\bm r}\uparrow} +
b^{\dagger}_{{\bm r}\downarrow} b_{{\bm r}\downarrow}
+ b^{\dagger}_{{\bm r}0} b_{{\bm r}0} -1 ).
\end{eqnarray}

 The spin operators $S^{\mu}_{\bm r}$ are bilinear forms
  of Schwinger bosons
\begin{subequations}\label{sop}
\begin{align}
S^{z}_{\bm r} &= {\bm b}^{\dagger}_{{\bm r}} {\cal S}^z {\bm b}^{\;}_{{\bm r}}=
b^{\dagger}_{{\bm r}\uparrow} b^{\;}_{{\bm r}\uparrow} - b^{\dagger}_{{\bm
r}\downarrow} b^{\;}_{{\bm r}\downarrow}, \\
S^{+}_{\bm r} &= {\bm b}^{\dagger}_{{\bm r}} {\cal S}^+ {\bm b}^{\;}_{{\bm r}}=
\sqrt{2} (b^{\dagger}_{{\bm r}\uparrow} b^{\;}_{{\bm r}0} +
b^{\dagger}_{{\bm r}0} b^{\;}_{{\bm r}\downarrow}), \\
S^{-}_{\bm r} &= {\bm b}^{\dagger}_{{\bm r}} {\cal S}^- {\bm b}^{\;}_{{\bm r}}=
\sqrt{2} (b^{\dagger}_{{\bm r}\downarrow} b^{\;}_{{\bm r}0} +
b^{\dagger}_{{\bm r}0} b^{\;}_{{\bm r}\uparrow}),
\end{align}
\end{subequations}
 where we defined
 \begin{equation}\label{Eq:spinor}
{\bm b}_{{\bm r}} =
\left( \begin{array} {ccc}
b_{{\bm r}\uparrow} & b_{{\bm r}0} &
b_{{\bm r}\downarrow}
\end{array} \right)^T.
\end{equation}

With these expressions, we can write ${\cal H}$ as
\begin{eqnarray}
\label{H2}
{\cal H} =   \sum_{{\bm r},\nu,\mu} J_{\mu} {\bm b}^{\dagger}_{\bm r} {\cal S}^{\mu} {\bm b}^{\;}_{{\bm r}}
{\bm b}^{\dagger}_{{\bm r}+{\bm e}_{\nu}} {\cal S}^{\mu} {\bm b}^{\;}_{{\bm r}+{\bm e}_{\nu}}
+D \sum_{{\bm r}} (1-{\bm b}^{\dagger}_{\bm r} {\cal A} {\bm b}^{\;}_{{\bm r}}),
\nonumber \\
\end{eqnarray}
where ${\cal A}_{\alpha,\beta}=\delta_{\alpha, 0}\delta_{\beta, 0}$.
The last term in this expression represents $(S^z_{{\bm r}})^2$ which
reduces to $1 - b^{\dagger}_{{\bm r} 0} b^{\;}_{{\bm r} 0}$ because
 of the constraint \eqref{constraint}. The product state at large $D$ is recovered if we introduce the condensate of $b_{{\bm r}0}$ boson, i.e., replace
$b_{{\bm r}0}$ and $b^{\dagger}_{{\bm r}0}$ operators by $b^{\dagger}_{{\bm r}0} = b^{\;}_{{\bm r}0} = s$ and set $s=1$.  By continuity,
$s$  remains nonzero in the whole paramagnetic state.

After condensing $b^\dagger_{{\bm r}0}$ in Eq.~\eqref{sop}, spin operators $S^{\pm}_{\bm r}$ become proportional to
$(b^{\dagger}_{{\bm r}\uparrow}  + b^{\;}_{{\bm r}\downarrow})$ while $S^{z}_{\bm r}$ retains its quadratic form \eqref{sop}.
The quadratic form of the spin-wave Hamiltonian \eqref{H2} can now be written easily as
\begin{eqnarray}
{\bar {\cal H}}_{sw} &=& \sum_{{\bm r},\nu,\sigma} s^2 J (b^{\dagger}_{{\bm r}\sigma} + b^{\;}_{{\bm r}{\bar \sigma}})
( b^{\dagger}_{{\bm r}+{\bm e}_{\nu} {\bar \sigma}} + b^{\;}_{{\bm r}+{\bm e}_{\nu} \sigma})
\nonumber \\
& &  + \mu \sum_{{\bm r}, \sigma} n_{{\bm r}\sigma}  + N (\mu-D) (s^2 -1) ,
\end{eqnarray}
where $\sigma=\{ \uparrow, \downarrow \}$, and $N$ is the total number of sites. The constraint is imposed on {\em average}, via the replacement
$\mu_{\bm r} \to \mu$. By Fourier transforming the bosonic operators,
$
{b}^{\;}_{{\bm k}\sigma} = \frac{1}{\sqrt{N}} \sum_{\bm r} b^{\;}_{{\bm r}\sigma} e^{i {\bm k}\cdot{\bm r}}
$, we
 obtain
 ${\bar {\cal H}}_{sw}$ in momentum space:
\begin{eqnarray}
{\bar {\cal H}}_{sw} &=& \sum_{{\bm k},\sigma}
(\mu + s^2 \epsilon_{\bm k} )  {b}^{\dagger}_{{\bm k}\sigma}  {b}^{\;}_{{\bm k}{ \sigma}}
+ N (\mu-D) (s^2 -1)
\nonumber \\
&+& \sum_{{\bm k},\sigma} \frac{ s^2 \epsilon_{\bm k}}{2}
({b}^{\dagger}_{{\bm k}\sigma}  {b}^{\dagger}_{-{\bm k}{\bar \sigma}}+\text{H.c.}),
\end{eqnarray}
with $\epsilon_{\bm k}= 2 J \gamma_{\bm k}$, $\gamma_{\bm k}=\sum_{\nu}  \cos{{\bm k} \cdot {\bm e}_{\nu}}$.
${\bar {\cal H}}_{sw}$ is diagonalized by means of a Bogoliubov transformation:
\begin{equation}
{b}^{\;}_{{\bm k}\sigma} = u_{\bm k} {\gamma}^{\;}_{{\bm k}\sigma}  + v_{\bm k} {\gamma}^{\dagger}_{-{\bm k}{\bar \sigma}},
\label{Bogo}
\end{equation}
with
\begin{subequations}
\begin{align}
u_{\bm{k}} &= (\mu + \omega_{\bm{k}})/ \left(2\sqrt{\mu \omega_{\bm{k}}}\right),\\
v_{\bm{k}} &= (\mu - \omega_{\bm{k}})/ \left(2\sqrt{\mu \omega_{\bm{k}}}\right),\\
\omega_{\bm k} &= \sqrt{\mu^2 + 2 \mu s^2 \epsilon_{\bm k}}.
\end{align}
\label{nnn}
\end{subequations}

The diagonal form of ${\bar {\cal H}}_{sw}$ is:
\begin{equation}\label{Eq:diagonal_form}
{\bar {\cal H}}_{sw} \! =\! N (\mu-D) (s^2 -1) \! + \! \sum_{{\bm k}\sigma} \! \left[ \omega_{\bm k}
(\gamma^{\dagger}_{{\bm k}\sigma} \gamma^{\;}_{{\bm k}\sigma} + \frac{1}{2}) - \frac{\mu}{2} \right] \! .
\end{equation}

\begin{figure}[!tbp]
	\centering
	\includegraphics[width=\columnwidth]{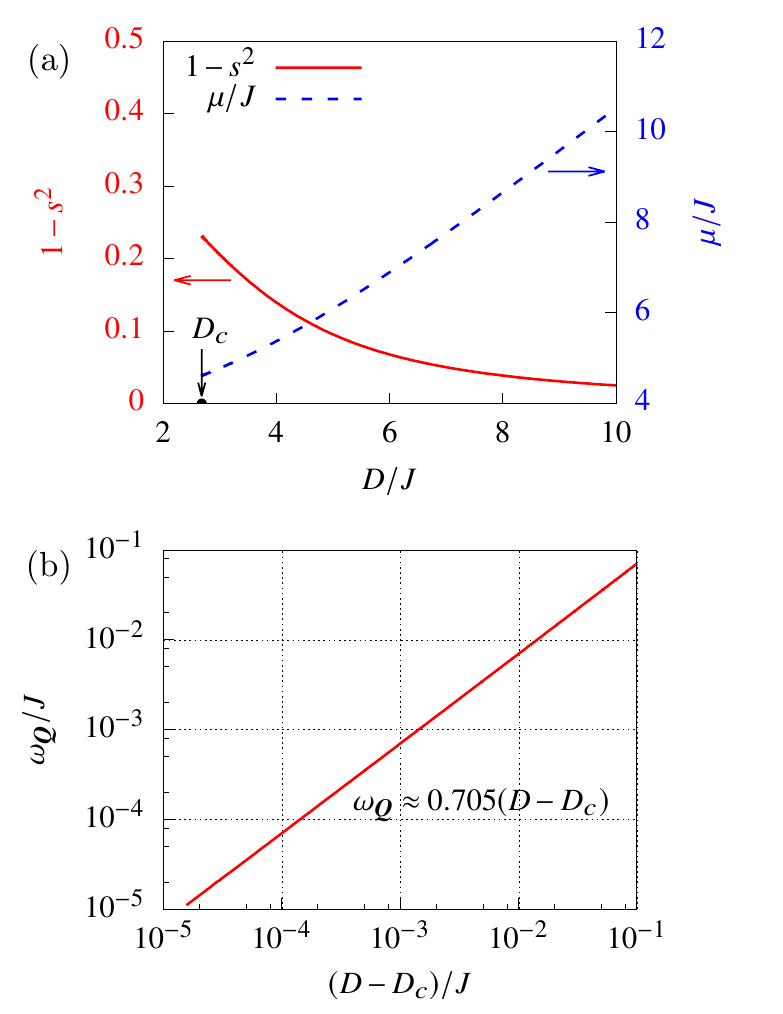}
	\caption{(a) Dependence of saddle-point parameters $s$ and $\mu$ on $D/J$
		[as indicated in Eq.~\eqref{Eq:self-consistent}, there is no $\zeta$ dependence]. $D_c/J \approx 2.68$ indicates the quantum critical point of single-magnon condensation
	below which long-range magnetic order develops. (b) Magnitude of the single-magnon gap $\omega_{\bm Q}$ near the critical point $D-D_c \ll J$. }
	\label{fig:s-mu}
\end{figure}

The dispersion relation $\omega_{\bm k}$ has minima at the wave vectors $\pm{\bm Q}
=\pm (4\pi/3,0)$,
  and the paramagnetic state remains stable against spin-wave excitations as long as $\omega^2_{\bm Q} >0$.
The variational parameters $s$ and $\mu$ are obtained from the saddle-point equations,
$\partial E_0^\text{PM} / \partial s   =0$ and $\partial E_0^\text{PM}/ \partial \mu  =0$, where $E_0^\text{PM}$ is the ground-state energy density:
\begin{equation}\label{Eq:energy_PM}
E_0^\text{PM} = \frac{1}{N} \sum_{\bm{k}} \omega_{\bm{k}} + (\mu-D) (s^2 -1) -\mu.
\end{equation}
The resulting self-consistent equations are:
\begin{subequations}\label{Eq:self-consistent}
\begin{align}
s^2 &= 2 -  \int \frac{d\bm{k}}{V_\text{BZ}} \frac{\mu + s^2 \epsilon_{\bm k}}{\omega_{\bm k} }, \\
D &= \mu \left (  1 +  \int \frac{d\bm{k}}{V_\text{BZ}} \frac{\epsilon_{\bm k}}{\omega_{\bm k} } \right ),
\end{align}
\end{subequations}
where $V_\text{BZ} = \frac{8\pi^2}{\sqrt{3}}$ is the size of the first Brillouin zone.

The single-magnon gap  $\omega_{\bm Q}$ vanishes at
 the phase boundary between the paramagnetic and the magnetically ordered phases.
 Combining this condition with Eq.~\eqref{Eq:self-consistent} we
obtain
 the critical value $D_c \approx 2.68J$ [see Fig.~\ref{fig:s-mu}].
The downward renormalization of $D_c$ from its naive single-particle value of $3 J$ to $D_c \approx 2.68J$ is caused by the renormalization
of the large-$D$ paramagnetic ground state by quantum fluctuations, which are captured by our mean-field parameters $s$ and $\mu$.
 Note that the value of $D_c$, obtained this way, is in much better
 agreement with numerical results~\cite{Zhang13},  than $D_c = 6J$, obtained using  more traditional Holstein-Primakoff--type  approach~\cite{Holstein40}
  (see Appendix B).

\subsection{Interaction between modes}
\label{sec:interaction}

 To analyze two-magnon bound states in a many-body system, we have to know the interaction between magnons.
It
 comes from the
Ising part of the Heisenberg interaction:
\begin{equation}
\label{int1}
{\cal H}^{(4)}_{\rm I} = \zeta J \sum_{{\bm r},\nu}  (n_{{\bm r}\uparrow} -  n_{{\bm r}{\downarrow}})
( n_{{\bm r}+{\bm e}_{\nu} {\uparrow}} -  n_{{\bm r}+{\bm e}_\nu \downarrow}).
\end{equation}
The signs of separate terms in (\ref{int1}) show that the interaction is
 repulsive between magnons of the same spin and attractive between magnons with opposite spins.
    In momentum space
\begin{equation}
\label{int}
{\cal H}^{(4)}_{\rm I} \!=\!  \frac{1}{N} \!\! \sum_{{\bm k}_1, {\bm k}_2, {\bm q},\sigma,\sigma^\prime} \!\!\!\! V_{\sigma \sigma'}({\bm q}) b^{\dagger}_{\bm{k}_1+\bm{q} \sigma} b^{\dagger}_{\bm{k}_2-\bm{q} \sigma'}
b_{\bm{k}_2 \sigma^\prime} b_{\bm{k}_1 \sigma},
\end{equation}
with
$
V_{\sigma \sigma^\prime}({\bm q}) = \sigma \sigma^\prime \zeta J \gamma_{\bm q}.
$

We will show below that an attractive interaction between the $\uparrow$ and $\downarrow$ magnons induces a two-particle bound state with $S_z=0$
in the full many-body system.
The
 energy $E$ of this bound state
 vanishes at a critical value of $D=D^b_c > D_c$ {\em above} the single-magnon condensation transition,
 like in the earlier analysis of a single exciton.
A vanishing gap of a two-magnon bound state signals a divergence
 of the susceptibility of an  order parameter
 which is bilinear in spin operators. Based on our previous discussion, the obvious candidate
   is
    vector chirality. Condition
     $D^b_c > D_c$ means
      that the chiral susceptibility diverges
       while the ordinary
       magnetic
        susceptibility is still finite. This implies
       the quantum paramagnetic state
        and the magnetically ordered state  are separated by the intermediate CL state.
        Crucial for this consideration is the fact that the single-magnon spectrum is two-fold degenerate,
with minima at $\pm {\bm Q}$. This gives two choices for the sign of vector chirality and in CL state the system chooses one particular sign, spontaneously breaking $Z_2$ chiral symmetry.

To see this, we expand ${\cal H}^{(4)}_{\rm I}$ in terms of the Bogoliubov quasi-particle operators \eqref{Bogo}
as
\begin{align}\label{Eq:interactions}
{\cal H}^{(4)}_{\rm I} &= \frac{1}{N} \!\! \sum_{\bm{k}_1,\bm{k}_2,\bm{q}} \!\! V_{\bm{q}}^{22o} (\bm{k}_{1},\bm{k}_{2})\gamma_{\bm{k}_{1}+\bm{q}\uparrow}^{\dagger}\gamma_{\bm{k}_{2}-\bm{q}\downarrow}^{\dagger}\gamma_{\bm{k}_{2}\downarrow}\gamma_{\bm{k}_{1}\uparrow} \nonumber \\
&+ \frac{1}{N} \!\! \sum_{\bm{k}_1,\bm{k}_2,\bm{q}, \sigma} \!\! V_{\bm{q}}^{22s} (\bm{k}_{1},\bm{k}_{2})\gamma_{\bm{k}_{1}+\bm{q}\sigma}^{\dagger}\gamma_{\bm{k}_{2}-\bm{q}\sigma}^{\dagger}\gamma_{\bm{k}_{2}\sigma}\gamma_{\bm{k}_{1}\sigma} \nonumber \\
&+ \frac{1}{N} \!\! \sum_{\bm{k}_1,\bm{k}_2,\bm{q}, \sigma} \!\! \Big[ V_{\bm{q}}^{31} (\bm{k}_1,\bm{k}_2) \gamma_{\bm{k}_1+\bm{q}\sigma}^\dagger \gamma_{\bm{k}_1 \sigma} \gamma_{\bm{k}_2 \sigma} \gamma_{-\bm{k}_2+\bm{q}\bar{\sigma}} \nonumber \\
&+ V_{\bm{q}}^{40} (\bm{k}_1,\bm{k}_2) \gamma_{-\bm{k}_{1}-\bm{q}\uparrow}\gamma_{-\bm{k}_{2}+\bm{q}\downarrow}\gamma_{\bm{k}_{2}\downarrow}\gamma_{\bm{k}_{1}\uparrow} + \text{H.c.} \Big].
\end{align}
 The  interaction vertices between spin-up and
 -down particles are given by
\begin{subequations}\label{Eq:interactions_detail}
\begin{align}
V_{\bm q}^{22o} ({\bm k}_1,{\bm k}_2) &= 2 \Big[ V_{\uparrow \uparrow}({\bm{k}_1 + \bm{k}_2})
B_{{\bm k}_1, {\bm k}_2} B_{{\bm{k}_1+\bm{q}}, {\bm{k}_2-\bm{q}}}
\nonumber \\
&\quad + V_{\uparrow \downarrow}({\bm q}) A_{{\bm k}_1, \bm{k}_1+{\bm q}}  A_{{\bm k}_2, \bm{k}_2 -{\bm q}}    \Big], \\
V_{\bm q}^{22s} ({\bm k}_1,{\bm k}_2) &=
 \frac{V_{\uparrow \uparrow}({\bm k}_1 - {\bm k}_2 + {\bm q})}{2} A_{\bm{k}_2, \bm{k}_1+\bm{q}} A_{\bm{k}_1, \bm{k}_2-\bm{q}}   \nonumber \\
&\quad + \frac{V_{\uparrow \uparrow}({\bm q})}{2} A_{{\bm k}_1, \bm{k}_1+{\bm q}}  A_{{\bm k}_2, \bm{k}_2 -{\bm q}} ,\\
V_{\bm q}^{31} ({\bm k}_1,{\bm k}_2) &= V_{\uparrow\uparrow}(\bm{k}_{1}-\bm{k}_{2}+\bm{q})A_{\bm{k}_{2},\bm{k}_{1}+\bm{q}}B_{\bm{k}_{1},\bm{k}_{2}-\bm{q}} \nonumber \\
&\quad + V_{\uparrow\uparrow}(\bm{q})A_{\bm{k}_{1},\bm{k}_{1}+\bm{q}}B_{\bm{k}_{2},\bm{k}_{2}-\bm{q}}, \\
V_{\bm q}^{40} ({\bm k}_1,{\bm k}_2) &= -\frac{V_{\uparrow\uparrow}(\bm{k}_{1}-\bm{k}_{2}+\bm{q})}{2}B_{\bm{k}_{1},\bm{k}_{2}-\bm{q}}B_{\bm{k}_{2},\bm{k}_{1}+\bm{q}} \nonumber \\
&\quad + \frac{V_{\uparrow\uparrow}(\bm{k}_{1}+\bm{k}_{2})}{2}B_{\bm{k}_{1},\bm{k}_{2}}B_{\bm{k}_{1}+\bm{q},\bm{k}_{2}-\bm{q}},\label{Eq:interaction_40}
\end{align}
\end{subequations}
where
\begin{subequations}\label{Eq:AB}
\begin{align}
 A_{{\bm k}_1, {\bm k}_2} &\equiv  u_{{\bm k}_1} u_{{\bm k}_2} - v_{{\bm k}_1} v_{{\bm k}_2} = \frac{\omega_{\bm{k}_{1}}+\omega_{\bm{k}_{2}}}{2\sqrt{\omega_{\bm{k}_{1}}\omega_{\bm{k}_{2}}}},\\
 B_{{\bm k}_1, {\bm k}_2} &\equiv  u_{{\bm k}_1} v_{{\bm k}_2} - v_{{\bm k}_1} u_{{\bm k}_2} = \frac{\omega_{\bm{k}_{1}}-\omega_{\bm{k}_{2}}}{2\sqrt{\omega_{\bm{k}_{1}}\omega_{\bm{k}_{2}}}}.
\end{align}
\end{subequations}
And, we remind, $V_{\sigma \sigma^\prime}({\bm q}) = \sigma \sigma^\prime \zeta J \gamma_{\bm q}$.

\subsection{Order Parameter and its Equation of Motion}
\label{sec:eom}

It is useful to develop some intuition before
addressing the issue of bound-state condensation in the
 many-body problem represented by the Hamiltonian \eqref{Eq:interactions}.
In
   the analysis in Sec.~\ref{sec:exciton} we selected {\em staggered vector chirality}  as the candidate for the two-magnon order parameter.
   (Neighboring up- and down-pointing triangles have  opposite chirality.)
This order parameter is expressed as
\begin{equation}
\kappa = \frac{1}{N} \sum_{{\bm r} \in \triangledown} \hat{z}\cdot \sum_{j=1}^3  \langle {\bm S}_{\bm r} \times {\bm S}_{{\bm r} + {\bm e}_j} \rangle,
\label{eq:kappa1}
\end{equation}
where the sum is over down-pointing triangles of the lattice and
  we go
  clockwise within a triangle.
The brackets denote average over the ground state. In momentum space
the vector chirality reads
\begin{equation}
\kappa = -\frac{1}{N} \sum_{\bm q} \sum_{j=1}^3 \sin[{\bm q} \cdot {\bm e}_j] \langle S^+_{\bm q} S^{-}_{-{\bm q}} \rangle.
\label{eq:kappa2}
\end{equation}
In terms of Bogoliubov eigenmodes, this becomes
\begin{eqnarray}
\kappa &=& -\frac{2s^2}{N} \sum_{\bm q} \sum_{j=1}^3 \sin[{\bm q} \cdot {\bm e}_j] (u_{\bm q} + v_{\bm q})^2  \nonumber\\
&&\times \langle (\gamma^\dagger_{-{\bm q} \uparrow} + \gamma_{{\bm q} \downarrow})
(\gamma_{-{\bm q} \uparrow} + \gamma^\dagger_{{\bm q} \downarrow}) \rangle .
\label{eq:kappa3}
\end{eqnarray}
In the low-energy long-wavelength
 approximation we focus on the lowest-energy magnons with ${\bm q} = \pm {\bm Q}$.
In terms of these magnons, vector chirality
is expressed as
\begin{equation}
\kappa
 = - \frac{3\sqrt{3} \mu s^2}{N \omega_{\bm Q}} \langle \gamma^\dagger_{-{\bm Q} \uparrow} \gamma^\dagger_{{\bm Q} \downarrow} -
\gamma^\dagger_{{\bm Q} \uparrow} \gamma^\dagger_{-{\bm Q} \downarrow}  + {\text{H.c.}}\rangle.
\label{eq:kappa4}
\end{equation}
This
 result shows that
  vector
   chirality is associated with the appearance of the bound state in the antisymmetric $S^z=0$ channel:
$\kappa$ changes sign under ${\bm Q} \to -{\bm Q}$ and is formed by $\uparrow$ and $\downarrow$ magnons.
 We then  introduce, by analogy with superconductivity,
  composite pair operators
\begin{subequations}\label{Eq:pair_operator}
	\begin{align}
	\phi_{L}(\bm{k})&\equiv\gamma_{\bm{Q}-\bm{k}\uparrow}\gamma_{\bar{\bm{Q}}+\bm{k}\downarrow},\\
	\phi_{R}(\bm{k})&\equiv\gamma_{\bar{\bm{Q}}+\bm{k}\uparrow}\gamma_{\bm{Q}-\bm{k}\downarrow}.
	\end{align}
\end{subequations}
where here and below we label ${\bar {\bm Q}} \equiv - {\bm Q}$.  The long-wavelength limit corresponds to small ${\bm k}$.
 The vector chirality is related to average values of these pair operators
\begin{equation}
\kappa =  - \frac{3\sqrt{3} \mu s^2}{N \omega_{\bm Q}} \left(\phi^*_R + \phi_R - \phi^*_L - \phi_L\right).
 \end{equation}

\begin{figure}[tb]
	\centering
	\includegraphics[width=0.95\columnwidth]{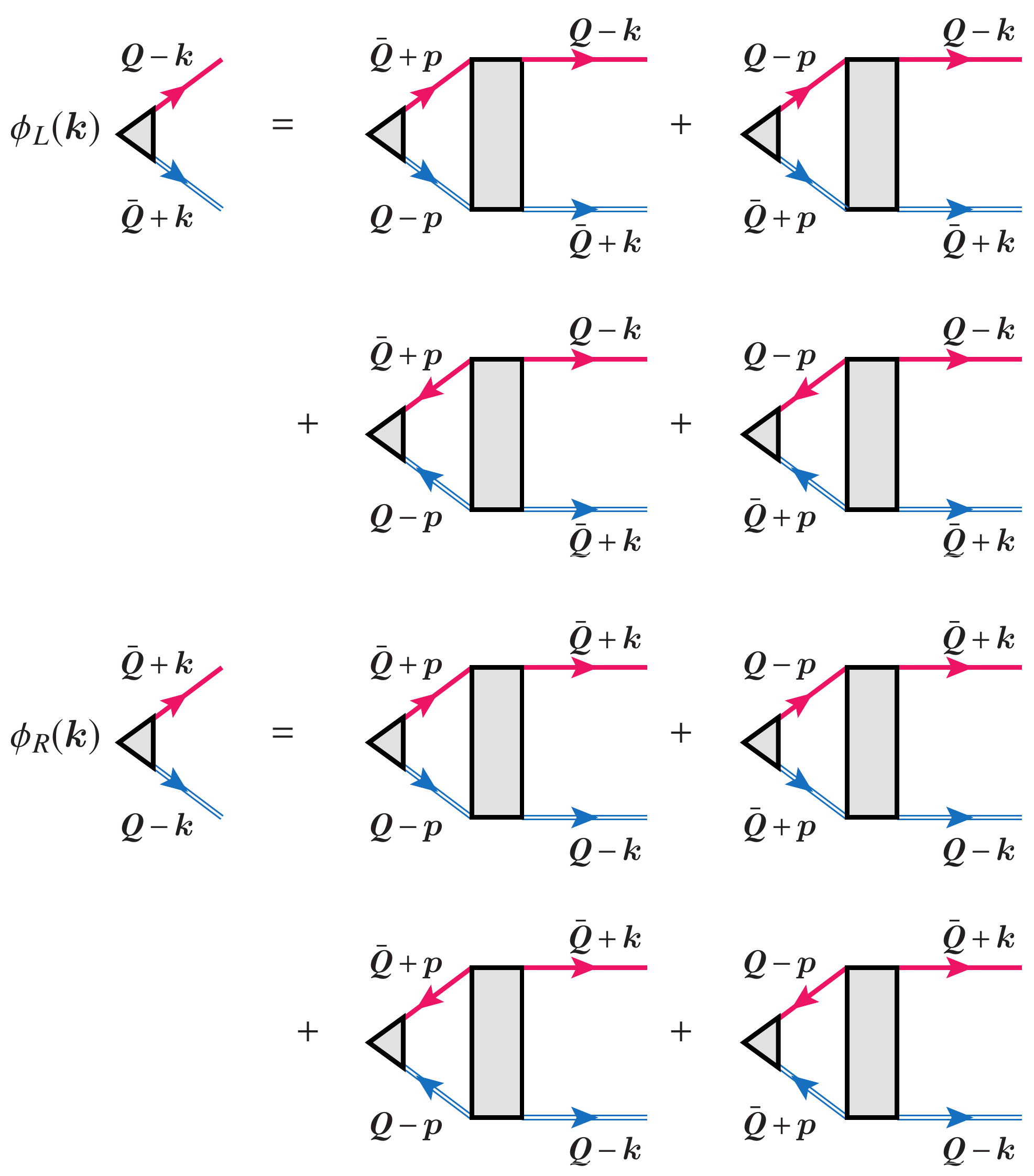}
\caption{Equations for
 the  vertices $\phi_{L/R}(\bm{k})$. The shaded
		rectangles denote the fully dressed irreducible  interactions between low-energy magnons.
 Solid (red) and hollow (blue) oriented lines represent spin $\uparrow$ and $\downarrow$ magnons, correspondingly.}
	\label{Fig:EOM}
\end{figure}

The equations for $\phi_{R/L}({\bm k})$
are presented graphically in Fig. \ref{Fig:EOM}.  The shaded vertices in this figure are fully dressed {\it irreducible} interactions in the particle-particle channel
(internal magnons have opposite frequencies $\pm \epsilon$ and momenta ${\bm Q} -{\bm p}$ and ${\bar {\bm Q}}+ {\bm p}$).
We verified that the particle number non-conserving dressed vertex $V_{\bm{q}}^{31}$
as well as the dressed vertex $V_{\bm{q}}^{22s}$, which is symmetric in the spin index $\sigma$,
   do not directly contribute to   the renormalizations of $\phi_{R/L}$.
 We also verified that magnon self-energy does not affect the formation of two-magnon bound state in any qualitative way and is therefore irrelevant for our purposes.
 Finally, the set of equations for $\phi_R$ and $\phi_L$ can be re-arranged as the subset for $\phi_R - \phi_L$ and the one for  $\phi_R + \phi_L$.
 We find that the pairing interaction is stronger for  $\phi_R - \phi_L$, in agreement with vector-chiral nature \eqref{eq:kappa4} of the anticipated order, and focus on it below
 (see \cite{Chubukov2013} for similar manipulations). 
  Collecting the diagrams in Fig.~\ref{Fig:EOM}, we obtain an integral equation
\begin{equation}
\label{Eq:EOM3}
\frac{1}{N} \!\! \sum_{\bm{p}} \frac{  F_{\bm{k},\bm{p}}^{22o} \, \theta_{\bm{p}}
- 4	F_{\bm{k},\bm{p}}^{04} \, \theta_{\bm{p}}^*}{2 \omega_{\bm{Q}-\bm{p}}}
= -  \theta_{\bm{k}}.
\end{equation}
Here, $\theta_{\bm{k}} \equiv 2 \omega_{{\bm Q}- {\bm k}} \langle \phi_R (\bm{k}) - \phi_L (\bm{k}) \rangle$ and 
\begin{subequations}
\begin{align}
F_{\bm{k},\bm{p}}^{22o} &\equiv \Gamma_{\bm{k}-\bm{p}}^{22o}(\bar{\bm{Q}} \! +\! \bm{p},\bm{Q} \! -\! \bm{p}) -
\Gamma_{2\bm{Q}-\bm{p}-\bm{k}}^{22o}(\bar{\bm{Q}} \! +\! \bm{p},\bm{Q} \! -\! \bm{p}), \\
F_{\bm{k},\bm{p}}^{04} &\equiv \Gamma_{\bm{k}-\bm{p}}^{04}(\bar{\bm{Q}} \! +\! \bm{p},\bm{Q} \! -\! \bm{p}) \! - \!
\Gamma_{2\bm{Q}-\bm{p}-\bm{k}}^{04}(\bar{\bm{Q}} \! +\! \bm{p},\bm{Q} \! -\! \bm{p}),
\end{align}
\end{subequations}
 where
 $\Gamma$'s are the fully dressed irreducible vertices between magnons with opposite frequencies.  Each $\Gamma$ term originates from  the corresponding interaction term in the Hamiltonian,
 e.g., $\Gamma_{\bm{k}-\bm{p}}^{22o}(\bar{\bm{Q}} \! +\! \bm{p},\bm{Q} \! -\! \bm{p})$ originates from   $V_{\bm{k}-\bm{p}}^{22o}(\bar{\bm{Q}} \! +\! \bm{p},\bm{Q} \! -\! \bm{p})$.
The factor $1/(2 \omega_{\bm{Q}-\bm{p}})$ comes from the integration over the frequency of internal magnon lines, e.g.
 \begin{equation}
 \int \frac{d\epsilon}{2\pi} \frac{1}{\epsilon - \omega_{\bm{Q}-\bm{p}} + i\delta} \frac{1}{-\epsilon - \omega_{-\bm{Q}+\bm{p}} + i\delta} = \frac{i}{2 \omega_{\bm{Q}-\bm{p}}}.
 \end{equation}

The equation
identical to \eqref{Eq:EOM3}
can be 
 also obtained 
 from the equation of motion for the chiral combination
$\theta_{\bm{k}}$, see \cite{Chubukov2013}.
The appearance of  a nontrivial  solution of  Eq. \eqref{Eq:EOM3}
 signals the instability of many-body paramagnetic ground state towards
 the condensation of two-magnon bound pairs.

Equation~\eqref{Eq:EOM3} highlights the role of the particle number non-conserving terms $
\Gamma_{\bm{q}}^{40}$ and $
\Gamma_{\bm{q}}^{04}$ [second line of \eqref{Eq:EOM3}].
Without them, the equation for $\theta_{\bm{k}}$ is U(1) degenerate: a solution $\theta_0$ is defined up to
a complex phase, i.e., there is a degeneracy in the order-parameter manifold.
 Such degeneracy is lifted by the anomalous terms, which depend on $\theta^*_{\bm p}$. The solutions with real and imaginary $\theta_{\bm p}$ now become different and the system chooses one of them.
One can easily verify that the state with  VC order parameter $\kappa \neq 0$ develops if the solution $\theta_p$ is real. The real solution preserves time-reversal symmetry,
as expected for the VC order. On the contrary,
if the solution of \eqref{Eq:EOM3} is imaginary, it yields
an order parameter that breaks
  time-reversal symmetry
   rather than vector chirality.

\subsection{Solution for the bound state at small $J_z$}
\label{sec:eom_1}
We now  analyze the structure of the interactions in  \eqref{Eq:EOM3} in the perturbative limit of small $J_z$.
Because each interaction term in the Hamiltonian has $J_z$ as the overall factor, a non-zero solution for $\theta_p$ emerges only if the overall smallness of the interaction is compensated by the singularity of the momentum integral in the kernel, much like it happens in BCS theory of superconductivity.  We argue below that the same happens in our case, but the singularity emerges at order $J^2_z$, once we include the renormalizations of the interaction vertices.   In this respect, the
 pairing that we find is similar to Kohn-Luttinger effect in the theory of superconductivity~\cite{kohn1965}.

\subsubsection{First order in $J_z$}

To first order in $J_z$, the vertices $\Gamma$ in Eq.~\eqref{Eq:EOM3} coincide with the interaction terms $V$ in the Hamiltonian.
There is a potential for singular behavior of the kernel as it contains  $1/\omega_{\bm{Q}-\bm{p}}$ which becomes singular at  $\bm{p} = 0$ and at  $D=D_c$, where single-magnon excitations condense.
One can easily check that at small ${\bm p}$ and small ${\bm k}$,
the prefactors for  $\theta_{\bm{p}}$ and $\theta_{\bm{p}}^*$ in ~\eqref{Eq:EOM3} are {\it negative}. This implies that (i) the pairing interaction is attractive, and (ii) the strongest attraction is for real $\theta_{{\bm k}}$.
 Using the
 explicit forms of the bare interactions [Eq.~\eqref{Eq:interactions_detail}], we find that at small ${\bm k}$ and ${\bm p}$ all four interactions $V$  in
Eq. ~\eqref{Eq:EOM3}
 are of order one in units of $J_z$, because
 $A_{\bm{k}_1,\bm{k}_2}$ and $B_{\bm{k}_1,\bm{k}_2}$ in (\ref{Eq:AB}) are $O(1)$:
 \begin{equation}
F_{\bm{k},\bm{p}}^{22o}
 =-4 F_{\bm{k},\bm{p}}^{04}
\approx-\frac{9}{4}J_z \frac{(\omega_{\bm{Q}-\bm{p}}+\omega_{\bm{Q}-\bm{k}})^{2}}{\omega_{\bm{Q}-\bm{p}}\omega_{\bm{Q}-\bm{k}}}.
\end{equation}
In this situation,  $\theta_{\bm{k}}$ depends on ${\bm k}$ in a non-singular way, and
 the condition that $\theta_{\bm k}$ is non-zero reduces to
\begin{equation}
1 =a \frac{J_z}{N} \sum_{\bm p} \frac{
1}{\omega_{\bm{Q}-\bm{p}}}.
\label{eq:eom4}
\end{equation}
where $a = O(1)$
 is a numerical coefficient.
 Since $\omega_{\bm{Q}}$ vanishes at $D=D_c$,  the kernel is singular. However, the singularity is integrable because $\omega_{\bm{Q}-\bm{p}}$ scales linearly in $|{\bm p}|$ at $D=D_c$.
 This implies that
there is no instability towards CL state at small $J_z$, as long as we use  bare interactions in Eq.~\eqref{Eq:EOM3}.

The reason for the  absence of
 the instability is related to specific property of $A_{\bm{k}_1,\bm{k}_2}$ and $B_{\bm{k}_1,\bm{k}_2}$ which determine the interaction terms $V$ in the Hamiltonian. Namely,
 when  $\bm{k}_1$ and $\bm{k}_2$ are  close to $\pm {\bm Q}$,
   $A_{\bm{k}_1,\bm{k}_2}$ and $B_{\bm{k}_1,\bm{k}_2}$ are $O(1)$. This is what we used in the derivation of Eq. (\ref{eq:eom4}).
On the other hand, if only one wave vector, say ${\bm k}_1$,  is near $\pm {\bm Q}$, i.e., $\omega_{{\bm k}_1}$ is small,
 while the other one, ${\bm k}_2$, is sufficiently far from $\pm {\bm Q}$ so that $\omega_{{\bm k}_2} = O(J)$,
  both $A_{\bm{k}_1,\bm{k}_2}$ and $B_{\bm{k}_1,\bm{k}_2}$ scale as
$\sqrt{\omega_{\bm{k}_2}/\omega_{\bm{k}_1}} \gg 1$. This implies that
 the interactions $V_{\bm{q}}^{22o}(\bm{k}_1,\bm{k}_2)$ and $V_{\bm{q}}^{40}(\bm{k}_1,\bm{k}_2)$ are
  $\sim O(1)$
only when all incoming/outgoing momenta are small, but
become much larger when one momentum remains near $\pm \bm{Q}$, while another one moves away from $\pm \bm{Q}$.

\begin{figure}[!btp]
	\centering
	\includegraphics[width=0.9\columnwidth]{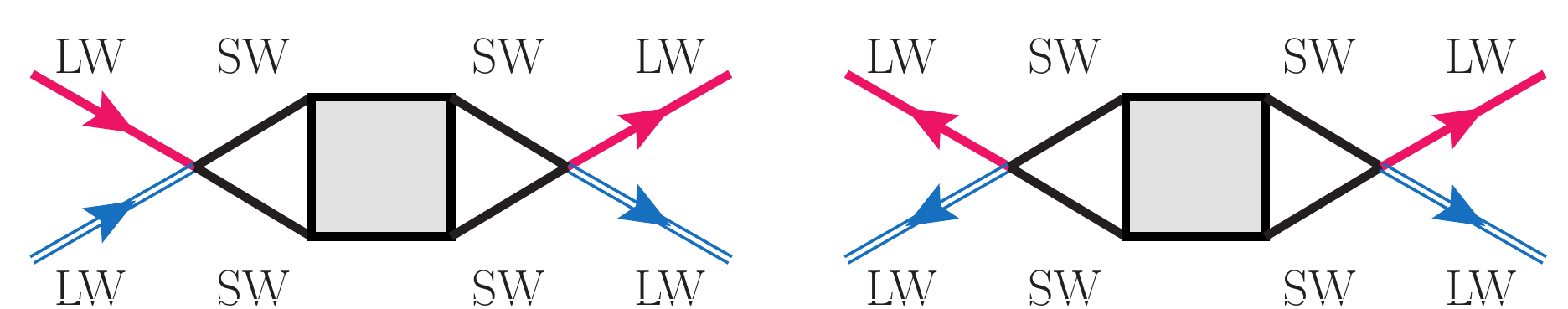}
	\caption{Schematics of the processes
 which contribute to relevant renormalization of the interactions $V_{\bm{q}}^{22o}$ (left) and $V_{\bm{q}}^{04}$ (right) between low-wavelength (LW) magnons by short-wavelength (SW)
 magnons with momenta far away from $\pm \bm{Q}$.  Complete list of diagrams to
	second order in $J_z$ is in  Figs.~\ref{Fig:2nd_normal} and \ref{Fig:2nd_anomalous}.
 The renormalized interactions $\Gamma$
	scale as $O(J^2_z/\omega_{\bm{Q}}^2)$ and are much stronger than bare interactions $V \sim J_z$. }
	\label{Fig:LW-SW}
\end{figure}

This observation suggests that one can potentially get a much stronger dressed interaction between low-energy bosons, if one includes
the renormalization of interaction vertices $V^{\cdots}_{\bm{q}}(\bm{k}_1,\bm{k}_2)$ by virtual processes involving bosons with momenta far away from ${\pm \bm{Q}}$ 
(see Fig.~\ref{Fig:LW-SW} for schematic illustration).
  To verify this, we now compute the dressed vertices $\Gamma$ to order $J^2_z$.

\subsubsection{Second order in $J_z$}
\label{app:ZW}

The  irreducible interactions $\Gamma$ to order $J^2_z$
come from three sets of processes: the $2\rightarrow2$ processes that conserves number of bosons,
and  $0\rightarrow4$ ($4\rightarrow0$) and $1\rightarrow3$ ($3\rightarrow1$) processes that create or annihilate additional bosons.
The external momenta in the vertices are fixed  at $p,\,k\ll Q$,
while the internal ones are not assumed to be small, and are integrated over the first Brillouin zone.

\begin{figure}[tb]
	\centering
	\includegraphics[width=1\columnwidth]{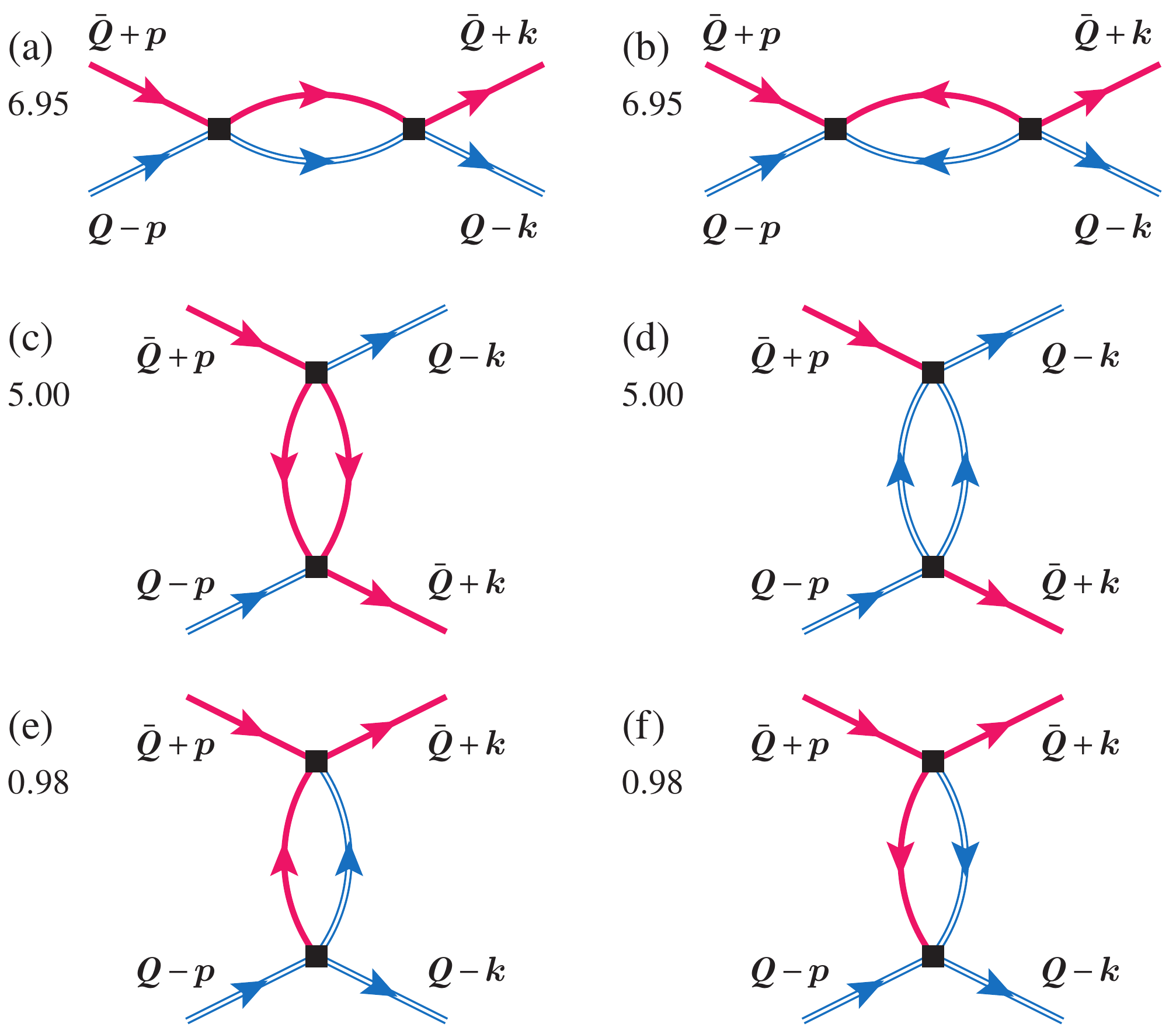}
	\caption{Nonzero second-order diagrams for the effective vertices $\Gamma_{\bm{k}-\bm{p}}^{22o}(\bar{\bm{Q}}+\bm{p},\bm{Q}-\bm{p})$.
	The diagrams for $\Gamma_{2\bm{Q}-\bm{k}-\bm{p}}^{22o}(\bar{\bm{Q}}+\bm{p},\bm{Q}-\bm{p})$
		are the same, except for different momentum labels. Solid (red) and hollow (blue) oriented lines represent Green's function of spin $\uparrow$ and $\downarrow$ magnons, correspondingly.
		The value of $\beta_i$ in
		Eq.~(\ref{Eq:normal_beta}) is listed separately for each diagram.}
	\label{Fig:2nd_normal}
\end{figure}

The relevant second order diagrams for $\Gamma_{\bm{k}-\bm{p}}^{22o}(\bar{\bm{Q}}+\bm{p},\bm{Q}-\bm{p})$
are shown in Fig.~\ref{Fig:2nd_normal}. The diagrams for $\Gamma_{2\bm{Q}-\bm{k}-\bm{p}}^{22o}(\bar{\bm{Q}}+\bm{p},\bm{Q}-\bm{p})$
are the same except for different momentum labels.  Summing up contributions from all six diagrams,
 we obtain
\begin{equation}\label{Eq:normal_beta}
F_{\bm{k},\bm{p}}^{22o}
= -\frac{\zeta^{2}s^{2}J^{3}}{8 \omega_{\bm{Q}-\bm{p}}\omega_{\bm{Q}-\bm{k}}}  \sum_{i=1}^6 \beta_i,
\end{equation}
where $\beta_i$  are numerical factors listed in
 Fig.~\ref{Fig:2nd_normal}.

\begin{figure}[tb]
		\includegraphics[width=0.95\columnwidth]{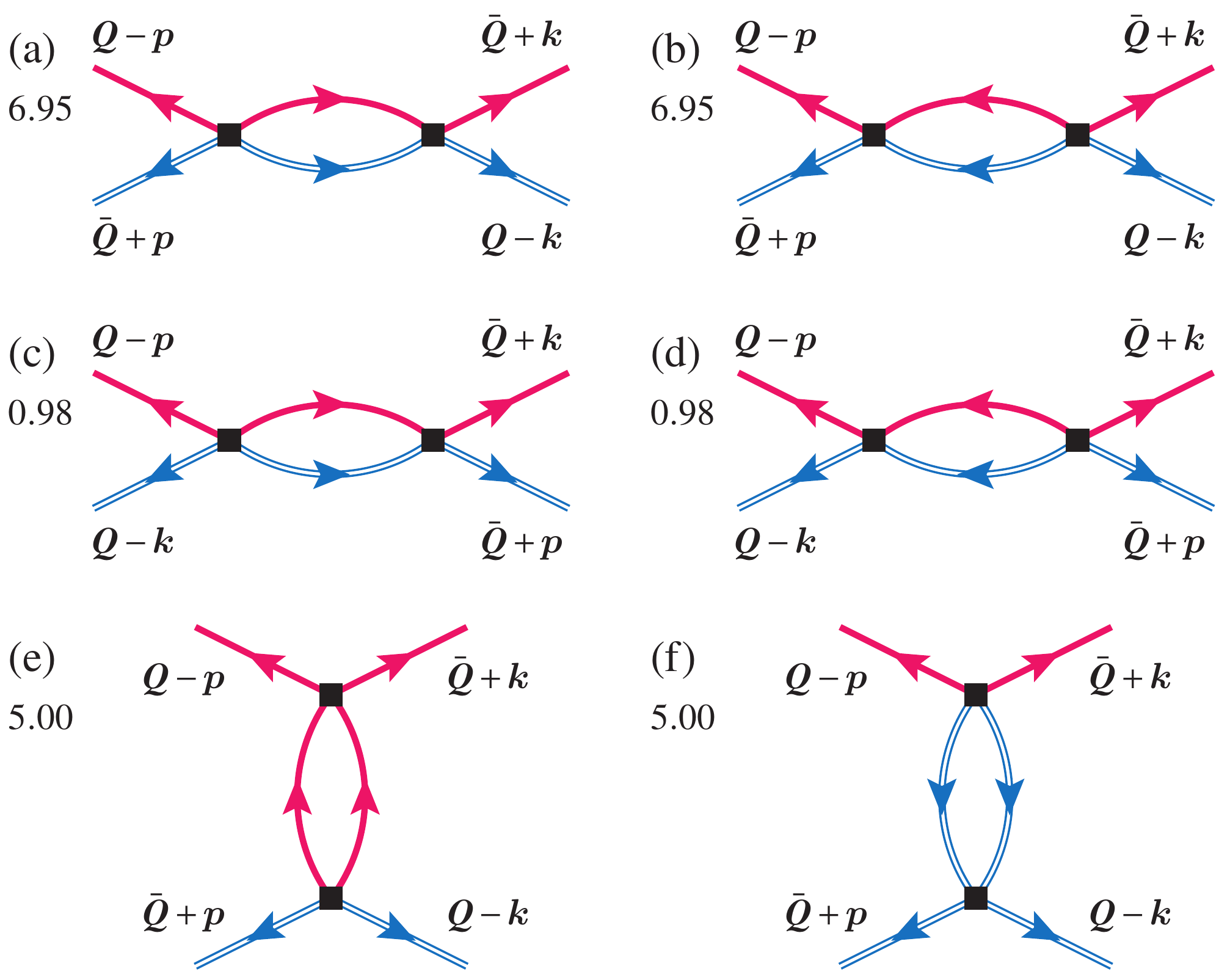}
		\caption{Nonzero second-order diagrams for the effective vertices $\Gamma_{\bm{k}-\bm{p}}^{04}(\bar{\bm{Q}}+\bm{p},\bm{Q}-\bm{p})$.
		The diagrams for $\Gamma_{2\bm{Q}-\bm{k}-\bm{p}}^{04}(\bar{\bm{Q}}+\bm{p},\bm{Q}-\bm{p})$
		are the same, except for different momentum labels. Solid (red) and hollow (blue) oriented lines represent Green's function of spin $\uparrow$ and $\downarrow$ magnons, correspondingly.
		The value of ${\bar \beta}_i$ in
		Eq.~(\ref{Eq:anomalous_beta}) is listed separately for each diagram.}
	\label{Fig:2nd_anomalous}
\end{figure}

Similarly, the second-order renormalization of the anomalous vertices
(see Fig.~\ref{Fig:2nd_anomalous}) yields
\footnote{One technical remark: a $1/2$ symmetrization factor should be included
when calculating diagrams in Figs.~\ref{Fig:2nd_normal}(c) and \ref{Fig:2nd_normal}(d) and
Figs.~\ref{Fig:2nd_anomalous}(e) and \ref{Fig:2nd_anomalous}(f), due to symmetrization of the
internal propagators.}
\begin{equation}\label{Eq:anomalous_beta}
-4 F_{\bm{k},\bm{p}}^{04}
= -\frac{\zeta^{2}s^{2}J^{3}}{8\omega_{\bm{Q}-\bm{p}}\omega_{\bm{Q}-\bm{k}}}  \sum_{i=1}^6 {\bar \beta}_i,
\end{equation}
where ${\bar \beta_i}$ are  numerical factors listed in Fig.~\ref{Fig:2nd_anomalous}.

We see that
\begin{align}
& \quad F_{\bm{k},\bm{p}}^{22o} =-4F_{\bm{k},\bm{p}}^{04} \nonumber \\
&  = -\frac{25.86 s^2}{8}~ \frac{\zeta^{2}J^{3}}{\omega_{\bm{Q}-\bm{p}}\omega_{\bm{Q}-\bm{k}}} \approx - 2.49 \frac{\zeta^{2}J^{3}}{\omega_{\bm{Q}-\bm{p}}\omega_{\bm{Q}-\bm{k}}},
\label{n_1_1}
\end{align}
where we used
 $s^2 = 0.77$ (at $D=D_c$) for the condensate of the
  $b_{\bm{r}0}$
   boson $s=\langle b_{{\bm r} 0}\rangle$ (see Fig.~\ref{fig:s-mu}).
We emphasize that (i) the dressed interaction  is negative, i.e., attractive, (ii) the interplay between normal and anomalous vertices remains exactly the same as
 for bare interaction, i.e., the largest attraction is for the real order parameter $\theta_{{\bm k}}$, and (iii) the attractive interaction now scales as
 $1/(\omega_{\bm{Q}-\bm{p}}\omega_{\bm{Q}-\bm{k}})$, i.e., the pairing vertex becomes truly singular at small $k$ and $p$.

Substituting Eq. (\ref{n_1_1}) into \eqref{Eq:EOM3} we obtain integral equation on $\theta_{{\bm k}}$, with $\alpha = 3.23 s^2 =2.49$:
\begin{equation}\label{Eq:EOM_eff}
\frac{1}{N}\sum_{\bm{p}}\frac{\alpha\zeta^{2}J^3}{\omega_{\bm{Q}-\bm{p}}^2 \omega_{\bm{Q}-\bm{k}} }\theta_{\bm{p}}
=\theta_{\bm{k}}.
\end{equation}
Equation~\eqref{Eq:EOM_eff} shows that the combination ${\cal C}= \omega_{\bm{Q}-\bm{k}} \theta_{\bm k}$ is actually ${\bm k}$ independent. This allows one to transform it into the self-consistent equation which reads as
\begin{equation}
\frac{1}{\alpha\zeta^{2} J^3} =
\frac{1}{N}\sum_{\bm{p}}\frac{1}{\omega^3_{\bm{Q}-\bm{p}}} \approx \frac{1}{N}\sum_{\bm{p}}\frac{1}{(\omega_{\bm{Q}}^{2}+9J^2 s^{4}p^{2})^{3/2}}.
\label{nnnn}
\end{equation}
The integral in the right-hand side is easily evaluated to be $1/(18 \pi J^2 s^4 \omega_{\bm Q})$. Importantly, it scales as
$1/\omega_{\bm{Q}}$ and therefore diverges as $D \to D_c$.
Using $\omega_{\bm Q} \approx 0.705 (D-D_c)$ (see Figure~\ref{fig:s-mu}), we obtain the critical value $D^b_c$ for the instability towards CL state:
\begin{equation}
D^b_c = D_c + 0.042 \alpha \zeta^2 J
\label{eq:scaling}
\end{equation}
 We see that $D^c_b > D_c$ for arbitrary small $\zeta= J_z/J$, hence, there is no threshold on the strength of the interaction for the emergence of CL phase.
\footnote{Note  that the scaling \eqref{eq:scaling} is different from
the one in Ref.~\cite{Chubukov2013}, due to the different asymptotic behavior of the interactions in that model;
in Ref.~\cite{Chubukov2013} the interaction scales as $1/\omega_{\bm{Q}}^2$ already at the bare level.}

\begin{figure*}[!tbp]
	\includegraphics[width=0.9\textwidth]{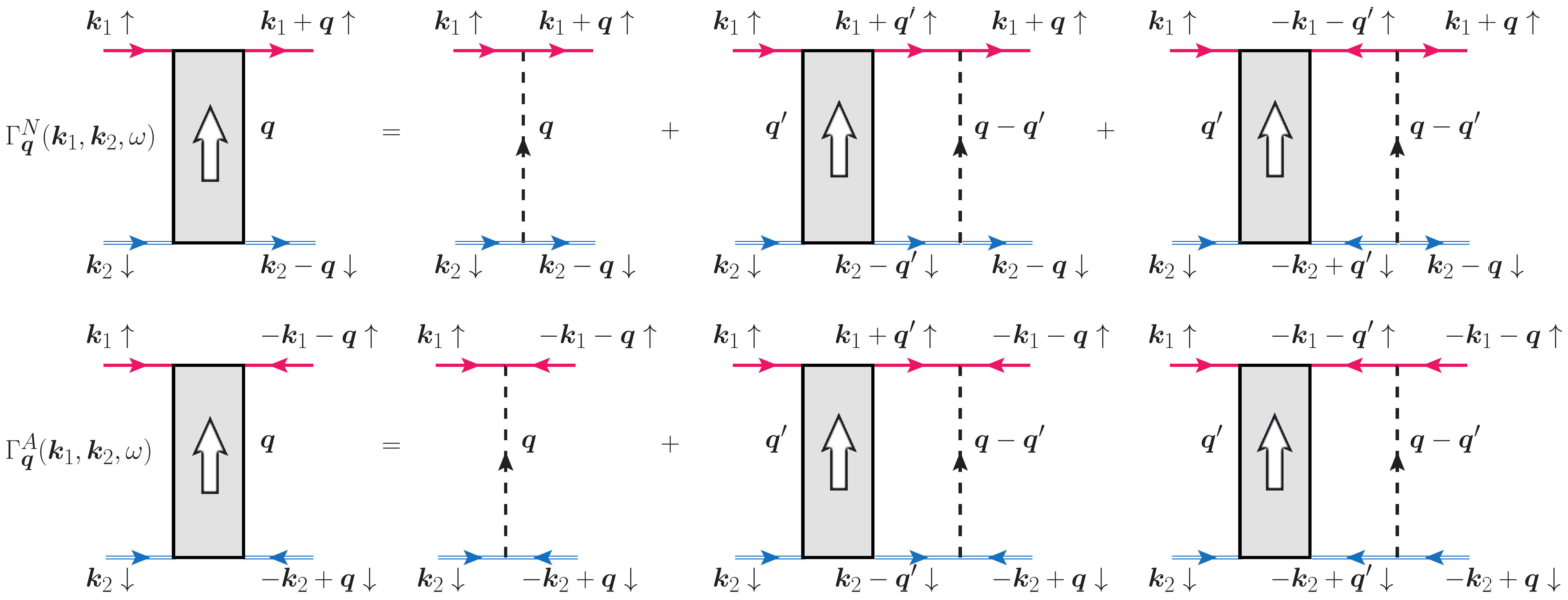}
	\caption{The diagrammatic representation of the Bethe-Salpeter equation. }
	\label{Fig:ladder}
\end{figure*}

Before we move to the analysis at arbitrary $J_z$, a comment is in order.   Within the saddle-point approximation of Eq.~\eqref{Eq:self-consistent},  $\omega_{\bm Q} \propto (D-D_c)$
 [we used this relation
 above
  to obtain  \eqref{eq:scaling}].
 This  approximation  becomes exact in the limit  $N \to \infty$,
 where $N$ is the number of bosonic flavors~\cite{Zhang13}.
  In contrast, the more traditional Holstein-Primakoff mean-field
approach leads to $\omega_{\bm Q} \propto (D-D_c)^{1/2}$
 (see Appendix).
None of these exponents is actually the exact one
because the dimension of the effective theory
in our case, $D_{\rm dim}=d+1= 3$, is lower than the upper critical
dimension $D_{\rm dim}=4$.
Moreover, a
 perturbative ($\epsilon$-expansion) renormalization group analysis shows that there is no stable fixed point for the {\it simultaneous} breaking of the
continuous U(1) and the discrete $Z_2$ symmetries~\cite{Kamiya2011,Parker2017}, i.e.,
at $J_z=0$  the transition at $D_c$ would be
 weakly first order.
If we take this into account, we find that the intermediate CL phase still emerges, but  for $J_z$ above some
 small but finite  value.
This is because $\omega_{\bm Q}$ remains
finite at a first-order transition, and a finite $J_z$ is needed for the bound state to form.

\subsection{Bethe-Salpeter equation}
\label{sec:BS}

In the previous section, we
obtained the instability of a paramagnet towards a CL state at small $J_z$ by analyzing the equations on the pair fields
$\phi_{L}(\bm{k})$ and $\phi_{R}(\bm{k})$.
 In this section, we use a complementary approach and extract the
  information about two-particle bound states
 from the poles of the
 four-point
  vertex function.
 This last approach can be rigorously justified in the opposite limit when $J_z$ is large enough such that the instability towards CL state occurs  while the density of bosons is still small.   The bosonic density is $\langle b^\dagger_{{\bm k} \sigma} b_{{\bm k} \sigma} \rangle = v_{\bm k}^2$, where, we remind,   $v_{\bm k}$ is the Bogoliubov parameter,  defined in Eqs. (\ref{Bogo}) and (\ref{nnn}). Below we assume that $v_{\bm k}$ is small at $D = D^b_c$ and keep only the leading-order terms in $v_{\bm k}$. In our notations, $v_{\bm k}$ is small when the Lagrange multiplier $\mu$ is large [see Eq.~\eqref{nnn}].  Figure \ref{fig:s-mu} shows that $\mu$ is large in a wide range of $D > D_c$.

  The fully renormalized normal and anomalous four-point vertex functions  $\Gamma_{\bm q}^N (\bm{k}_1, \bm{k}_2, \omega)$ and $\Gamma_{\bm{q}}^{A}(\bm{k}_{1},\bm{k}_{2},\omega)$,  with incoming frequency $\omega$,   are obtained
 by solving the
Bethe-Salpeter
(BS)
equations.  Within our approximation, these equations reduce to the ones  shown in  Fig.~\ref{Fig:ladder}.
 In analytic form \footnote{Observe that incoming/outgoing lines in Fig.~\ref{Fig:ladder} belong to particles with opposite spin indices $\sigma$.
 As a result, there is no need to symmetrize BS
equations. This explains the absence of usual factors of $\frac{1}{2}$ in front of the interaction terms in \eqref{Eq:interactions}.}
\begin{subequations}\label{Eq:Bethe-Salpeter}
\begin{align}
&\quad \Gamma_{\bm q}^N (\bm{k}_1, \bm{k}_2, \omega) - V_{\bm q}^{22o} (\bm{k}_1, \bm{k}_2)\nonumber \\
&= - \int \frac{d {\bm q}^\prime}{V_{\text{BZ}}} \frac{\Gamma_{\bm q^\prime}^N (\bm{k}_1, \bm{k}_2,\omega) V_{{\bm q}-{\bm q^\prime}}^{22o} ({\bm k}_1 +{\bm q}^\prime,{\bm k}_2 - {\bm q}^\prime)}{\omega_{{\bm k}_1+{\bm q}^\prime}+\omega_{{\bm k}_2-{\bm q}^\prime}-\omega} \nonumber \\
&- \int\frac{d\bm{q}^{\prime}}{V_{\text{BZ}}}\frac{16 \Gamma_{\bm{q}^{\prime}}^{A}(\bm{k}_{1},\bm{k}_{2},\omega) V_{\bm{q}-\bm{q}^{\prime}}^{04}(\bm{k}_{1}+\bm{q}^{\prime},\bm{k}_{2}-\bm{q}^{\prime})}{\omega_{\bm{k}_{1}+\bm{q}^{\prime}}+\omega_{\bm{k}_{2}-\bm{q}^{\prime}}+\omega}, \\
&\quad \Gamma_{\bm{q}}^{A}(\bm{k}_{1},\bm{k}_{2},\omega) - V_{\bm{q}}^{40}(\bm{k}_{1},\bm{k}_{2}) \nonumber \\
&= -\int\frac{d\bm{q}^{\prime}}{V_{\text{BZ}}}\frac{\Gamma_{\bm{q}^{\prime}}^{N}(\bm{k}_{1},\bm{k}_{2},\Omega) V_{\bm{q}-\bm{q}^{\prime}}^{40}(\bm{k}_{1}+\bm{q}^{\prime},\bm{k}_{2}-\bm{q}^{\prime})}{\omega_{\bm{k}_{1}+\bm{q}^{\prime}}+\omega_{\bm{k}_{2}-\bm{q}^{\prime}}-\omega} \nonumber \\
& -\int\frac{d\bm{q}^{\prime}}{V_{\text{BZ}}}
\frac{\Gamma_{\bm{q}^{\prime}}^{A}(\bm{k}_{1},\bm{k}_{2},\Omega)V_{\bm{q}-\bm{q}^{\prime}}^{22o}
(\bm{k}_{1}+\bm{q}^{\prime},\bm{k}_{2}-\bm{q}^{\prime})}{\omega_{\bm{k}_{1}+\bm{q}^{\prime}}+\omega_{\bm{k}_{2}-\bm{q}^{\prime}}+\omega}.
\end{align}
\end{subequations}
  Note that this set does not contain the interaction $V^{31}$.
 According to Eqs. \eqref{Eq:interactions_detail} and \eqref{Eq:AB}, $V^{31}$
contains an
 additional factor of $v_{\bm k}$ and therefore is smaller than $V^{22}$ interaction.
However we still need to include $V^{40}$ and $V^{04}$ terms in the anomalous vertex, despite the fact that they contain
$v_{\bm k}^2$, because these terms fix the phase of the two-magnon order parameter, see discussion following \eqref{Eq:EOM3}.
 But even here,
$V^{31}$ vertices do not
 contribute to the renormalization of the
anomalous vertex  $\Gamma_{\bm q}^A$, again because they contain additional small factor of $v_{\bm k}$ compared to $V^{22}$ vertices.
 The second order diagrams ~(c) and (d) in Figs.~\ref{Fig:2nd_anomalous} are not included in the BS equation too.
 These terms are not relatively small in $v_{\bm k}$, however, given that they
  just reinforce the negative amplitude of $\Gamma_{\bm q}^A$, we do not expect these terms to give rise to any qualitative changes.

There are two special c.m. momenta: $\bm{K}=0$ ($\bm{k}_2=-\bm{k}_1=\bm{Q}$) and $\bm{K}=2\bm{Q}$ ($\bm{k}_2 = \bm{k}_1 = \bm{Q}$).
For each case, we fix the incoming momenta $\bm{k}_1$ and $\bm{k}_2$, and discretize the
 momentum
  $\bm{q}$ in the first BZ of the triangular lattice.
We then solve Eq. ~\eqref{Eq:Bethe-Salpeter} numerically.

Implementing this procedure,
we obtained
 that bound state appears at a finite frequency $\omega$  already
for arbitrarily small $\zeta$. This is an expected result because in 2D
the density of states has a  logarithmic  singularity   at the bottom of the magnon band.
 The appearance of the bound state should not be confused with the instability towards CL state. The latter occurs when the  frequency of the bound state reduces down to zero.

\begin{figure}[!tbp]
	\includegraphics[width=\columnwidth]{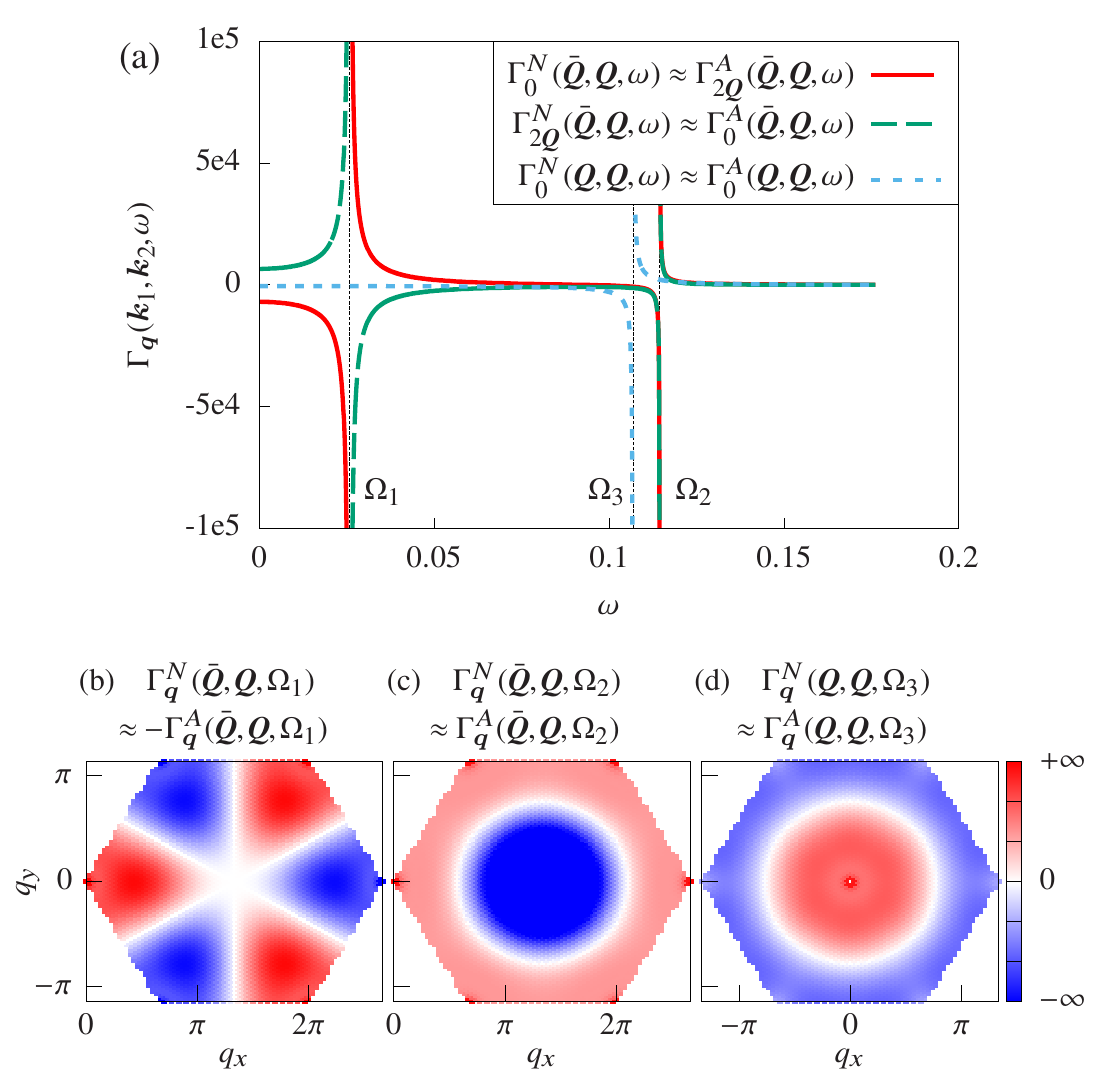}
	\caption{
		The four-point vertex function solved by discretizing Eq.~\eqref{Eq:Bethe-Salpeter} on $60\times 60$ uniform mesh. The parameters are fixed at $J=\zeta=1, D=2.808$.
		(a) Energy dependence of $\Gamma_{\bm{q}}$ in three dominant scattering channels. The positions of the poles are denoted as $\Omega_\nu$ in the figure.
		(b)--(d) Momentum dependence of $\Gamma_{\bm{q}}$ at the poles $\Omega_\nu + \epsilon$, where $\epsilon \rightarrow +0$.
	}
	\label{Fig:pole}
\end{figure}

In a close similarity with the analysis of a single
 two-spin exciton in Sec. \ref{sec:exciton}, we find two
  bound-state solutions for $\bm{K}=0$, at frequencies $\Omega_1$ and $\Omega_2$, and one solution for $\bm{K}=2\bm{Q}$, at frequency $\Omega_3$.   The solutions have
  the following symmetry properties
   of four-point vertices
  (see Fig.~\ref{Fig:pole}):
\begin{subequations}
	\begin{align}
	\Gamma_{\bm{Q}+\bm{q}}^{N/A} (\bar{\bm{Q}},\bm{Q},\Omega_1) &= - \Gamma_{\bm{Q}-\bm{q}}^{N/A} (\bar{\bm{Q}},\bm{Q},\Omega_1), {\text{odd}} \\
	\Gamma_{\bm{Q}+\bm{q}}^{N/A} (\bar{\bm{Q}},\bm{Q},\Omega_2) &= \quad  \Gamma_{\bm{Q}-\bm{q}}^{N/A} (\bar{\bm{Q}},\bm{Q},\Omega_2), {\text{even}}\\
	\Gamma_{\bm{q}}^{N/A} (\bm{Q},\bm{Q},\Omega_3) &= \quad  \Gamma_{-\bm{q}}^{N/A} (\bm{Q},\bm{Q},\Omega_3), {\text{even}}.
	\end{align}
\end{subequations}

The two-particle propagator near the pole $\omega = \Omega_\nu$ ($\nu =1,2,3$) has the form
~\cite{Nakanishi1969,Ueda2013}:
\begin{align}\label{eq:lehmann}
&\quad G^{(2)}(\omega,\bm{K}=\bm{k}_1+\bm{k}_2,\frac{\bm{k}_1-\bm{k}_2}{2},\frac{\bm{k}_1-\bm{k}_2}{2}+\bm{q}) \nonumber \\
&\approx \frac{\langle 0 | \gamma_{\bm{k}_1 +\bm{q} \uparrow} \gamma_{\bm{k}_2 - \bm{q} \downarrow} | \Psi^{(\nu)} \rangle \langle \Psi^{(\nu)} | \gamma_{\bm{k}_1  \uparrow}^\dagger \gamma_{\bm{k}_2\downarrow}^\dagger |0 \rangle}{\omega - \Omega_\nu} \nonumber \\
&= \frac{\Psi_{\bm{K}}^{(\nu)}(\frac{\bm{k}_1-\bm{k}_2 }{2}+ \bm{q}) \Psi_{\bm{K}}^{(\nu)*}(\frac{\bm{k}_1-\bm{k}_2}{2})}{\omega - \Omega_\nu},
\end{align}
where $\Psi_{\bm{K}}^{(\nu)}(\bm{k})$ is the two-particle wave function with total momentum $\bm{K}$ and relative momentum $\bm{k}$.

Alternatively, we can obtain this two-particle propagator from the self-energy corrections. Near the pole  at $\omega=\Omega_\nu$,
\begin{align}
\label{eq:propagator}
&\quad G^{(2)}(\omega,\bm{K},\frac{\bm{k}_1-\bm{k}_2}{2},\frac{\bm{k}_1-\bm{k}_2}{2}+\bm{q}) \nonumber \\
 &\approx
 G_0^{(2)}(\Omega_\nu,\bm{K},\frac{\bm{k}_1-\bm{k}_2}{2})
 \nonumber \\
& \quad \times \Gamma_{\bm q}^N (\bm{k}_1, \bm{k}_2, \Omega_\nu) \cdot G_0^{(2)}(\Omega_\nu,\bm{K},\frac{\bm{k}_1-\bm{k}_2}{2} + \bm{q}).
\end{align}

\begin{figure}[!tbp]
	\centering
	\includegraphics[width=\columnwidth]{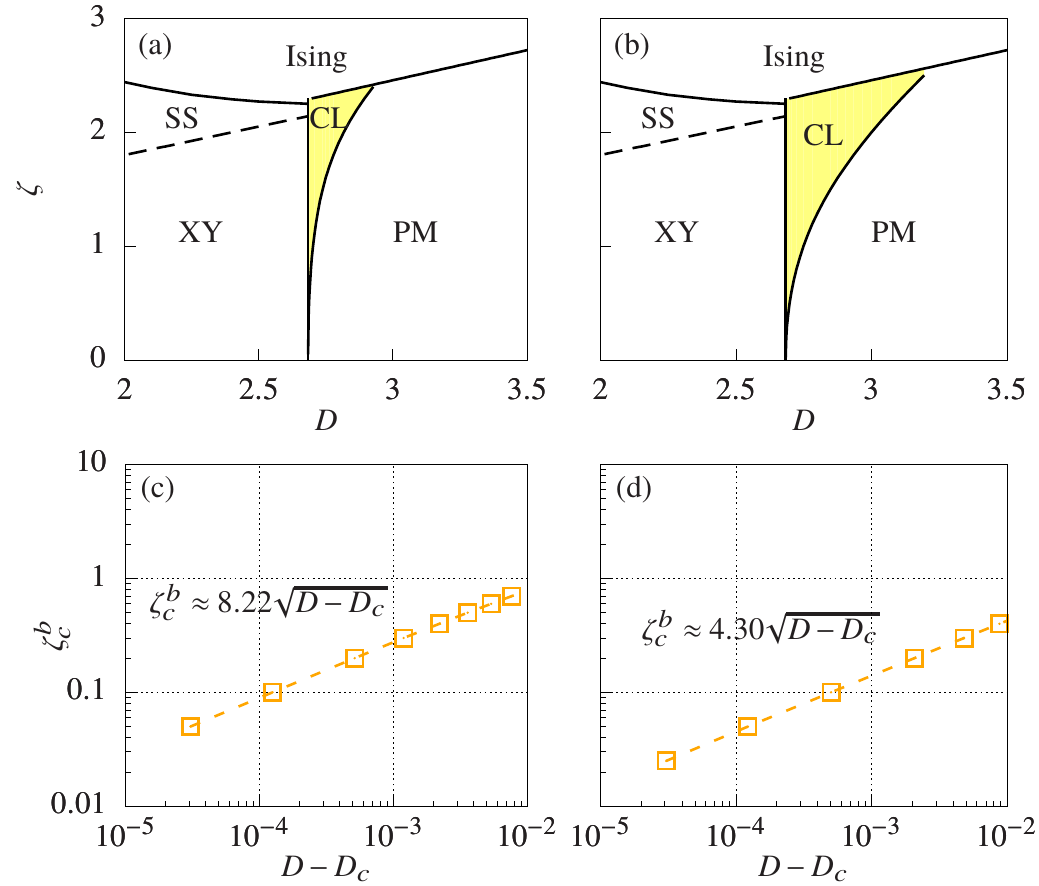}
	\caption{(a), (b): Phase diagram of the XXZ model with single-ion anisotropy Eq.~\eqref{eq:ham}.
				The CL/PM phase boundary is obtained by solving the BS equation Eq.~\eqref{Eq:Bethe-Salpeter} on a non-uniform mesh,
		where more points are sampled near the singular part of the interactions until convergence.
		The Ising/PM, Ising/SS, and SS/XY phase boundaries are discussed in Appendix~\ref{sec:Ising}.
		In (a) we used the solution of
		Eq.~\eqref{Eq:Bethe-Salpeter} with only normal interactions present.
In (b) we used the full solution of
 Eq.~\eqref{Eq:Bethe-Salpeter}. (c), (d) The chiral phase boundaries in log-log scale
 for panels (a) and (b), respectively.}
	\label{Fig:phd}
\end{figure}

Combining Eqs.~\eqref{eq:lehmann} and \eqref{eq:propagator}, we can connect the symmetry of
 four-point interaction vertices to the symmetry of two-particle wave functions:
\begin{subequations}
\begin{align}
\frac{\Gamma^N_{\bm{Q}+\bm{q}}(\bar{\bm{Q}},\bm{Q},\Omega_\nu)}{\Gamma^N_{\bm{Q}-\bm{q}}(\bar{\bm{Q}},\bm{Q},\Omega_\nu)} &=
\frac{\Psi_{\bm{K}=0}^{(\nu)}(\bm{q})}{\Psi_{\bm{K}=0}^{(\nu)}(-\bm{q})},\\
\frac{\Gamma^N_{\bm{q}}(\bm{Q},\bm{Q},\Omega_\nu)}{\Gamma^N_{-\bm{q}}(\bm{Q},\bm{Q},\Omega_\nu)} &=
\frac{\Psi_{\bm{K}=2\bm{Q}}^{(\nu)}(\bm{q})}{\Psi_{\bm{K}=2\bm{Q}}^{(\nu)}(-\bm{q})}.
\end{align}
\end{subequations}

From these relations we can
 extract the symmetry of the bound-state wave functions:
\begin{subequations}
	\begin{align}
	\label{nnnn_a}
	\Psi_{\bm{K}=0}^{(1)} (\bm{q})&= - \Psi_{\bm{K}=0}^{(1)} (-\bm{q}), \\
	\Psi_{\bm{K}=0}^{(2)} (\bm{q})&= \quad  \Psi_{\bm{K}=0}^{(2)} (-\bm{q}), \\
	\Psi_{\bm{K}=2\bm{Q}}^{(3)} (\bm{q})&= \quad \Psi_{\bm{K}=2\bm{Q}}^{(3)} (-\bm{q}).
	\end{align}
\end{subequations}
 We found that out of three bound-state frequencies, the smallest one is $\Omega_1$.  We see from (\ref{nnnn_a}) that
  the corresponding wavefunction is odd under spatial inversion,
  consistent with the symmetry of the chiral order parameter $\kappa$.

When $\zeta$ increases at a constant $D$, or $D$ decreases at a constant $\zeta$,
 the attractive interaction between bosons with opposite flavors
  also increases, and the bound-state frequency
  $\Omega_1$ decreases and eventually reaches zero.
The softening of the $\Omega_1$  mode signals the onset of the
  CL phase.

We show the location of the transition into the CL phase in Fig.~\ref{Fig:phd}.  The solid line between CL/PM in Fig.~\ref{Fig:phd}a shows the location of the boundary of the CL phase, obtained numerically by
 keeping in the BS equation ~\eqref{Eq:Bethe-Salpeter}  only the normal interaction $V^{22o}$ (i.e., only particle number conserving processes).
Figure~\ref{Fig:phd}b shows the location of the CL phase boundary  obtained  by solving the full Eq.~\eqref{Eq:Bethe-Salpeter}, keeping
 both $V^{22o}$ and $V^{40}, V^{04}$ interactions.
 In both cases, the  phase boundary
 is obtained by requiring that  the
 pole frequency is zero, $\Omega_1 =0$.


Although the analysis in this section is justified when $D^b_c$ is substantially
 larger
 than $D_c$, which requires
 $\zeta$ of order one, it is nevertheless useful to compare the results of this and the previous sections.
  In the
 previous Sec.~\ref{sec:eom_1},  we found  that the instability towards CL state at small $\zeta$ is related to singular behavior of the
  dressed pairing interaction, which scales as
   $
   \zeta^2/\omega_{\bm{Q}}^2$.
   A naive discretization of
Eq.~\eqref{Eq:Bethe-Salpeter} using a  uniform mesh of $210\times 210$ points in the first BZ does not capture the singular part of the interaction.
To obtain the
 boundary of the CL phase at small $\zeta$,
 we used a non-uniform mesh which is much denser near the singular region of Eq.~\eqref{Eq:Bethe-Salpeter}.
The phase diagrams shown in Fig.~\ref{Fig:phd}
 were verified by sampling $\sim 5000$
 points near $\bm{q}=\{0,2\bm{Q}\}$ on top of a $30 \times 30$ uniform background.

To further compare
 the results obtained to second order in $J_z$
with the ones obtained by solving BS
equation, we label by
 $\alpha$ the overall numerical factor from the second-order diagrams.
 The full second-order result, Eq.~\eqref{n_1_1}, gives $\alpha = 2.49$.
If instead
 we pick only normal forward scattering process
from forward scattering normal vertices
[diagram (a) in Fig.~\ref{Fig:2nd_normal}], we obtain $\alpha\approx  0.33$.
 If we added up  all ladder contributions
[diagrams (a) and (b) in Figs.~\ref{Fig:2nd_normal}
and \ref{Fig:2nd_anomalous}],
we
 would
  obtain larger $\alpha\approx 1.34$.
 Observe that larger $\alpha$  leads, at fixed $\zeta$, to
  larger
  critical $D^b_c$ [see Eq.~\eqref{eq:scaling}].
This
 is consistent with the
 results obtained by solving BS equation,
Fig.~\ref{Fig:phd}.
 We recall that if we use all
second order diagrams, we obtain an even
larger $\alpha\approx 2.49$. This
 means that using only ladder diagrams in
 the BS equation
 gives a conservative estimate of the
critical $D^b_c$ for the instability towards the CL state.
In Figs.~\ref{Fig:phd}(c) and \ref{Fig:phd}(d), the critical scaling of the chiral phase boundary is found to be $\zeta_c^b \sim \left( D_c^b -D_c \right)^{1/2}$,
again in agreement with the $J_z^2$ analysis in the previous section.

\section{DMRG calculation of the single- and two-magnon gaps}
\label{sec:dmrg}

To provide further evidence that the two-magnon bound-state gap $\Delta_b$ closes before closing the single-magnon gap $\Delta_s$ upon decreasing $D$, we perform density matrix renormalization group (DMRG) calculations on $6\times 6$ triangular lattice with periodic boundary condition~\footnote{We have also studied a $6\times 18$ lattice with cylindrical boundary conditions, the gaps are then extracted by sweeping the center of the cylinder. The results are qualitatively the same as the $6\times 6$ PBC ones shown in the main text.}.
$M=6000$ states were kept in the calculation, leading to truncation error $< 10^{-4}$ for all the data points presented here.
\begin{figure}[!tbp]
\centering
\includegraphics[width=0.45\textwidth]{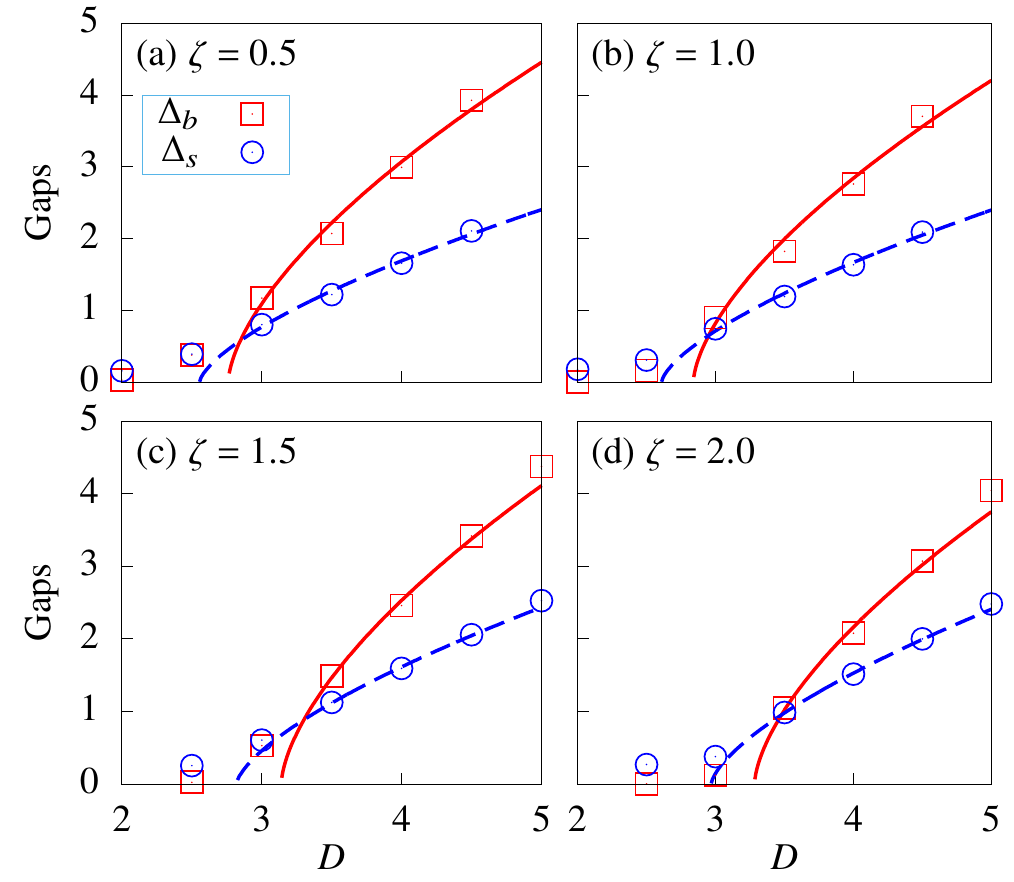}
\caption{Two-magnon gap $\Delta_b$ and single-magnon gap $\Delta_s$ obtained from DMRG on $6 \! \times \! 6$ triangular lattice with periodic boundary condition, with $J=1$. The solid (dashed) line represents fitting to the gap $\Delta_b$ ($\Delta_s$) by Eq.~\eqref{Eq:universality} with two parameters $\{c_b,\, D_b\}$  ($\{c_s,\, D_s\}$). Only data points before gap crossing are used in the fitting procedure.}
\label{Fig:dmrg}
\end{figure}

In DMRG, the two-magnon gap $\Delta_b$ (single-magnon gap $\Delta_s$) corresponds to the energy of the first excited state in the $S_z=0$ sector (ground state in the $S_z=1$ sector), measured from the $S_z=0$ ground state. Since the transition into the CL phase belongs to the $d=3$ Ising university class, the critical exponent $\nu$ is given by $\nu_{\text{Ising}} \approx 0.63$. Similarly, the transition into XY phase belongs to $d=3$ XY university class, giving $\nu_{\text{XY}}\approx 0.67$:
\begin{subequations}\label{Eq:universality}
\begin{align}
\Delta_b &= E_{S_z=0}^{(1)}-E_{S_z=0}^{(0)} = c_b (D-D_b)^{\nu_{\text{Ising}}}, \\
\Delta_s &= E_{S_z=1}^{(0)}-E_{S_z=0}^{(0)} = c_s (D-D_s)^{\nu_{\text{XY}}},
\end{align}
\end{subequations}
where the superscripts $(0)$ and $(1)$ denote the ground and the first excited state, respectively.

In Fig.~\ref{Fig:dmrg}, we calculate the evolution of the two gaps as a function of $D$, for four different values of $\zeta=\{0.5, 1.0, 1.5, 2.0 \}$. The data are then fitted to Eq.~\eqref{Eq:universality} with two fitting parameters (while keeping $\nu$ fixed to the known value). In all cases, the two gaps clearly cross each other before closing, indicating
 that the VC order emerges before single-particle excitations of the paramagnetic state soften.

As discussed in previous sections, the first excitation in the $S_z=0$ sector is odd under inversion,
  and its condensation signals the appearance of the VC order.
This can be checked numerically:
    by performing an
     exact diagonalization on $3\times 3$ and $3\times 6$ lattices, we can obtain the wave function of the lowest-energy states.
 The first excited state in the $S_z=0$ sector is always found to be odd under inversion.

\section{Conclusions}
\label{sec:summary}

In this work, we
 studied the sequence of quantum phase transitions  in the spin-$1$ triangular XXZ model,
induced by an  easy-plane single-ion anisotropy $D$ (see Fig.~\ref{fig:schematic}).
  Within non-interacting magnon approximation, the system is in a paramagnetic state at $D > D_c$ and in the XY-ordered phase at $D < D_c$.  We analyzed the effects of interactions and found that they change the phase diagram in a qualitative way.  Namely, we
 found that the  continuous U(1) symmetry and the discrete chiral Z$_2$ symmetry,
which are spontaneously broken in the XY ordered phase, break at different values $D$, implying the existence of an intermediate
chiral liquid phase in-between the XY and quantum paramagnetic phases. This liquid phase has  no magnetic ordering ($\langle \bm{S}_n \rangle =0$)  and is characterized by
a finite staggered vector chirality, $\langle { \kappa}_{nm} \cdot {\hat {\bm z}} \rangle \neq 0$, which has opposite sign on
 the neighboring
 triangles. It therefore spontaneously breaks spatial inversion symmetry. Note that the time-reversal symmetry is preserved. Our analytical results are supported by DMRG simulations on a $6\times 6$ triangular lattice.
Remarkably, we find the gapped chiral liquid phase to extend up to
large values of the exchange anisotropy ($\zeta > 2$), for which its window of stability reaches $D_c^b - D_c \simeq J/2$.
This rather large range of stability opens the possibility of observing this phase in real materials.

As we discussed in the Introduction, this is not the first time an Ising-type phase with nematic order parameter bilinear in microscopic spin degrees of freedom is observed.
However,
the previous and closely related observations~\cite{Chubukov2013,Parker2017}, involved semiclassical large spin ($S \gg 1$) expansion of Heisenberg models with pronounced spatial anisotropy subject to external magnetic field on triangular and kagom\'e lattices,
correspondingly. Our consideration is specific to a more `quantum' spin $S=1$ and spatially isotropic triangular lattice model and does not require an external magnetic field.

\begin{figure}[!tbp]
	\centering
	\includegraphics[width=0.8\columnwidth]{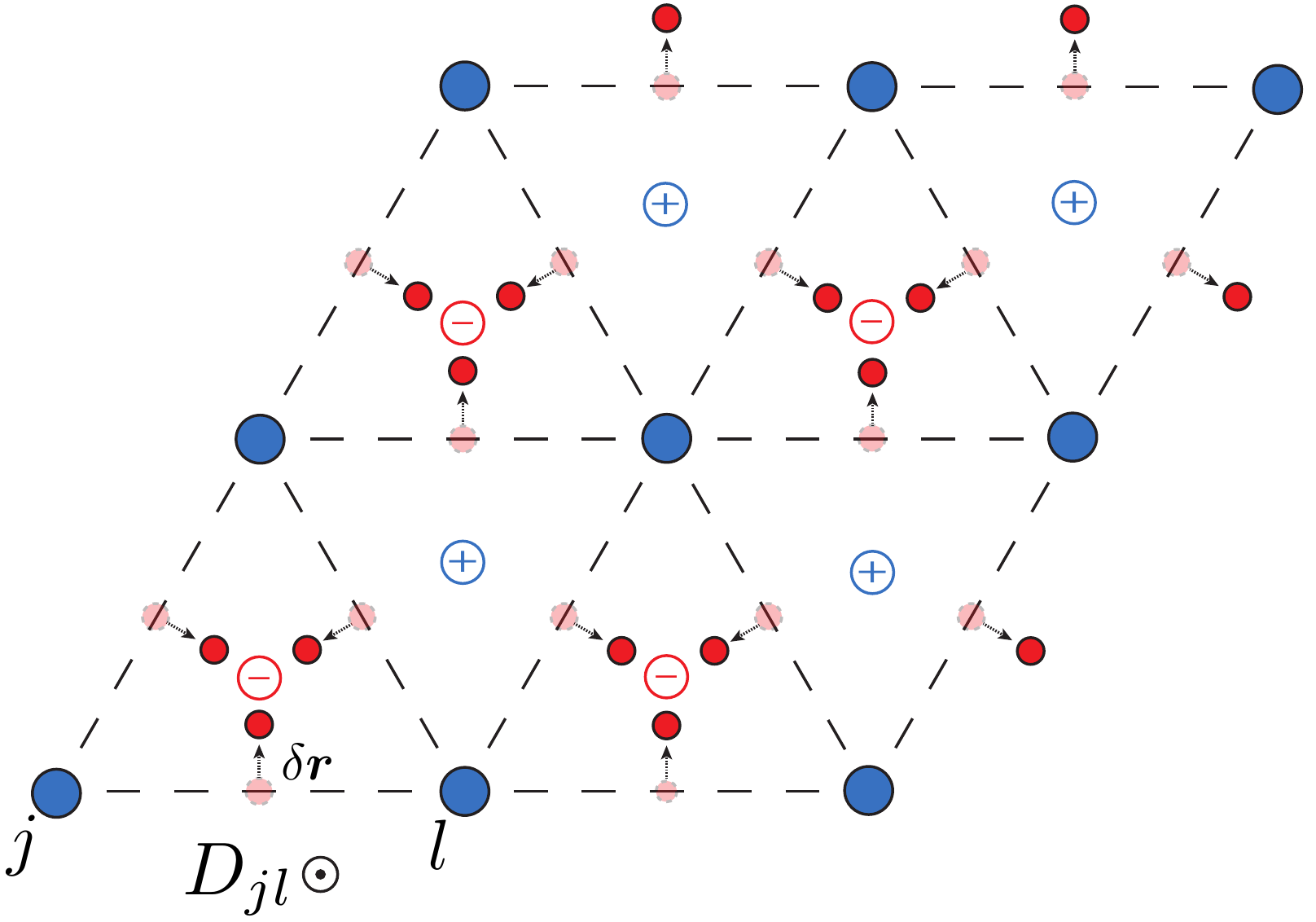}
	\caption{Schematic plot of the inverse Dzyaloshinskii-Moriya effect.}
	\label{Fig:inverse_DM}
\end{figure}

The experimental signatures of
the VC
order
have been discussed in Refs.~\cite{Chubukov2013,Parker2017}
 They are related to the  so-called ``inverse Dzyaloshinskii-Moriya (DM)''
 effect,
  which
  was proposed as a mechanism for multi-ferroic behavior of spiral magnets~\cite{Katsura05,PhysRevB.73.094434,PhysRevLett.96.067601}.
Namely, a local VC
order parameter $\langle \bm{S}_j \times \bm{S}_l \rangle$   produces a net electric dipole proportional to
$\bm{e}_{jl} \times \langle \bm{S}_{j}  \times  \bm{S}_{l}\rangle$
(where $\bm{e}_{jl} \equiv ({\bm r}_j - {\bm r}_l) / |{\bm r}_j - {\bm r}_l|$).
As shown in Fig.~\ref{Fig:inverse_DM}, the polarization is induced by the displacement $\delta \bm{r}$ of a medium ion (with charge $q_I$) away from the bond center.
This ion is typically an anion ($q_I < 0$),
 e.g., oxygen O$^{2-}$ for the case of transition metal oxides,
 and it mediates
the super-exchange interaction between spins $\bm{S}_j$ and $\bm{S}_l$.  The induced DM interaction, ${\bm{D}}_{jl} \propto \delta \bm{r} \times {\bm e}_{jl}$,
lowers the magnetic energy by
 $\bm{D}_{jl} \cdot \langle \bm{S}_{j}  \times  \bm{S}_{l}\rangle$,
which is linear in $\delta \bm{r}$. Because the elastic energy cost is quadratic in $\delta \bm{r}$, the local electric polarization $q_I \delta {\bm r}$
 becomes finite
 once  $\langle \bm{S}_{j}  \times   \bm{S}_{l}\rangle \neq 0$. As it is clear from Fig.~\ref{Fig:inverse_DM}, the ionic displacements
induced by the staggered
VC  ordering
lead to a charge density wave order, which can be detected with x rays.

It is worth stressing once again that {\em geometric frustration} is essential to our construction: spontaneous breaking of the inversion symmetry occurs via formation of the two-magnon bound state formed by
magnons at $\pm {\bm Q}$ which are degenerate in energy. Similar considerations apply to lattice models of strongly interacting bosons~\cite{Zaletel2014,Janzen2016,Zhu2016} with inverted (frustrated) sign
of particle's hopping between sites.
  There is a certain similarity between
   our results and
   loop current orders proposed
   for strongly correlated fermion models~\cite{Simon2002,Wang2014,Agterberg2015}.

From a statistical physics perspective, we can think of this quantum phase a transition as a classical phase transition in dimension $2+1$. It is known that the suppression of XY ordering is induced by proliferation of vortex lines that span the full system~\cite{Kleinert}. In the paramagnetic state, vortex and anti-vortices have the same probability of being at a given triangle. In the chiral liquid state, the vortices occupy one sublattice of triangles (e.g., the triangles that are pointing up) with higher probability, while anti-vortices occupy the other sublattice  (e.g., the triangles that are pointing down) with higher probability. In other words, the staggered chiral liquid is a vortex density wave.

Finally, it is important to note that our conclusions are far more general than the particular model that we have considered here. Our results suggest that exotic quantum liquid states are likely to emerge in the proximity of  quantum phase transitions between a $T=0$ paramagnet and quantum magnet that breaks both {\it continuous and discrete symmetries}. In other words, like in the case of metallic systems where quantum critical points guide the experimental search for unconventional superconductors and non-Fermi-liquid behavior, the quantum critical points of bosonic systems can play a similar role in the experimental search for exotic quantum liquids.

\section{Acknowledgements}
We thank Z. Nussinov, Y. Motome, and S. Zhang for helpful discussions.
The numerical results were obtained in part using the computational resources of the National
Energy Research Scientific Computing Center, which is supported by the Office of Science
of the U.S. Department of Energy under Contract No. DE-AC02-05CH11231.
Z.W. and C.D.B. are supported by funding from the Lincoln Chair of Excellence in Physics and
from the Los Alamos National Laboratory Directed Research and Development program.
O.A.S. is supported by the National Science Foundation Grant No. NSF DMR-1507054.
W.Z. is supported by DOE NNSA through LANL LDRD program.
A.V.C. is supported by Grant No. NSF DMR-1523036.

\appendix

\section{Other phases of the model}
\label{sec:Ising}

Here, we discuss the additional phases that appear in our model for  large enough $J_z$.
As shown in   Fig.~\ref{Fig:phd}, the PM and CL phases are bounded from above by an Ising-type {\em spin density wave} (SDW) phase, which becomes the ground state
for strong enough $\zeta J$. This phase, which  is described by the local SDW order parameter
$\langle S^z_{\bm r} \rangle \propto e^{i {\bm Q} \cdot {\bm r}}$,  corresponds to a three-sublattice ordering
with $\langle S^z_{\bm r} \rangle$ being positive in
one sublattice (A), negative in another sublattice (B), and equal to zero (disordered) on the third sublattice (C).
This  partially disordered AFM ordering is obtained for the triangular lattice $S=1$ Blume-Capel  model~\cite{Blume66,Capel66,Blume1971,Mahan78,Collins88}, which is obtained by setting  the $XY$ exchange to zero in our Hamiltonian ${\cal H}$ defined in Eq.~\eqref{eq:ham}.

At mean-field level, there is an intermediate phase preempting the transition between the Ising and XY phases.
This phase is characterized by coexistence of Ising-type SDW and in-plane magnetic ordering [which breaks U(1) symmetry] known as spin supersolid (SS) state~\cite{Penrose1956,Ng2006,Laflorencie2007,Sengupta2007,Sengupta2007_v2,Picon2008,Peters2009}. The SS phase is also a three-sublattice ordering, whose longitudinal components follow the same pattern as in the partially disordered Ising phase, while the transverse components form a collinear pattern
\begin{subequations}
\begin{align}
&S_{\bm{r}_A}^\perp = S_{\bm{r}_B}^\perp = - \alpha S_{\bm{r}_C}^\perp,\\
&S_{\bm{r}_A}^z = -S_{\bm{r}_B}^z \neq 0, ~~ S_{\bm{r}_C}^z = 0,
\end{align}
\end{subequations}
where $\alpha < 1$.  The collinear ordering is more favorable than the $120\degree$ structure because of the different  magnitudes of transverse spin components.

The boundary between the partially disordered Ising phase and the PM phase is determined by comparing their ground state energies. The ground state energy of the
PM phase is given in Eq.~\eqref{Eq:energy_PM}. The ground-state energy of the partially disordered Ising phase can be computed
by using the same Lagrange multiplier method that we applied to the quantum PM.

The spin operators are again represented by SU(3) Schwinger bosons in the fundamental representation [see Eqs.~\eqref{constraint}--\eqref{H2}].
The mean-field state of the Ising phase corresponds to condensation of different flavors of bosons in three sublattices:
\begin{equation}
b_{\bm{r}_A\uparrow}^{\dagger} = s_1,\quad
b_{\bm{r}_B\downarrow}^{\dagger} = s_{-1},\quad
b_{\bm{r}_C0}^{\dagger} = s_0.
\end{equation}
Due to time-reversal symmetry, we expect $s_1 = s_{-1}$, and $\mu_1 = \mu_{-1}$. The spin-wave Hamiltonian  is
\begin{align}
\bar{\mathcal{H}}_{sw} &= \sum_{\bm{k}} \Big[ \Psi_{\bm{k}}^\dagger \bar{H}_{sw} (\bm{k}) \Psi_{\bm{k}}  + (3 s_1^2  \zeta J + D + \mu_1 ) \nonumber \\
&\quad \qquad \times \left( b_{A\downarrow,\bm{k}}^\dagger b_{A\downarrow,\bm{k}} + b_{B\uparrow,\bm{k}}^\dagger b_{B\uparrow,\bm{k}} \right) \Big]  \nonumber \\
&\quad + \frac{N}{3} \Big[ (s_0^2-2)\mu_0 + (2s_1^2-3) \mu_1  \nonumber \\
&\quad \qquad  + (2s_1^2-1)D  -3s_1^4 \zeta J \Big],
\end{align}
where
\begin{equation}
\Psi_{\bm{k}} \equiv \begin{pmatrix}b_{A0,\bm{k}} & b_{C\downarrow,\bm{k}} & b_{B0,\bar{\bm{k}}}^{\dagger} & b_{C\uparrow,\bar{\bm{k}}}^{\dagger}\end{pmatrix}^{T},
\end{equation}
and the $4\times 4$ matrix $\bar{H}_{sw}(\bm{k})$ is:
\begin{equation}
\bar{H}_{sw}(\bm{k})
\! = \! \begin{pmatrix}
\mu_{1} & s_{0}s_{1} \Gamma_{\bar{\bm{k}}}& s_{1}^2 \Gamma_{\bm{k}}  & s_{0}s_{1} \Gamma_{\bar{\bm{k}}} \\
s_{0}s_{1} \Gamma_{\bm{k}} & \mu_{0}+D & s_{0}s_{1} \Gamma_{\bar{\bm{k}}} & 0 \\
s_{1}^2  \Gamma_{\bar{\bm{k}}} & s_{0}s_{1}  \Gamma_{\bm{k}} & \mu_{1} & s_{0}s_{1} \Gamma_{\bm{k}} \\
s_{0}s_{1} \Gamma_{\bm{k}} & 0 & s_{0}s_{1} \Gamma_{\bar{\bm{k}}} & \mu_{0}+D
\end{pmatrix},
\end{equation}
with $\Gamma_{\bm{k}} \equiv J \sum_\nu \exp (-i \bm{k} \cdot \bm{e}_\nu)$.

By diagonalizing the matrix $\text{diag}\{1,1,-1,-1\}\bar{H}_{sw}(\bm{k})$ we obtain the ground-state energy:
\begin{align}
E_{0}^\text{Ising}&=\frac{1}{N} \sum_{\bm{k}}\left(\omega_{+,\bm{k}}+\omega_{-,\bm{k}}\right) + \frac{1}{3} \Big[ (s_0^2-2)\mu_0  \nonumber \\
&\quad + (2s_1^2-3) \mu_1  + (2s_1^2-1)D  -3s_1^4 \zeta J \Big],
\end{align}
where
\begin{equation}
\omega_{\pm,\bm{k}}=\frac{1}{\sqrt{2}}\sqrt{\tau_{\bm{k}}\pm \sqrt{\kappa_{\bm{k}}}}.
\end{equation}
\begin{subequations}
	\begin{align}
	\tau_{\bm{k}}&=(D+\mu_{0})^2+\mu_{1}^{2}-s_{1}^{4} \left|\Gamma_{\bm{k}}\right|^{2},\\
	\kappa_{\bm{k}} &=  - 4 \mu_1^2 (D+\mu_0)^2 -4 s_1^2  (D+\mu_0)  \Big[ 2 s_0^2 s_1^2 \left( \Gamma_{\bm{k}}^3 + \Gamma_{\bar{\bm{k}}}^3 \right) \nonumber \\
	&\quad   -\left| \Gamma_{\bm{k}} \right|^2 \left( 4s_0^2 \mu_1 + s_1^2 (D+\mu_0) \right) \Big] +\tau_{\bm{k}}^2.
	\end{align}
\end{subequations}

The variational parameters $\{ s_0, s_1, \mu_0, \mu_1 \}$ are obtained from the saddle point equations
\begin{equation}
\frac{\partial E_0^\text{Ising}}{ \partial s_0} = 0, \,
\frac{\partial E_0^\text{Ising}}{ \partial s_1 } = 0, \,
\frac{\partial E_0^\text{Ising} }{ \partial \mu_0 }= 0,\,
\frac{\partial E_0^\text{Ising}}{ \partial \mu_1 }= 0.
\end{equation}

For large $\zeta$, the energy $E_0^\text{Ising}$ becomes lower than $E_0^\text{PM}$, corresponding to a first order phase transition between the quantum PM phase and the partially disordered Ising phase. The transition line, shown in Fig.~\ref{Fig:phd}, is determined by  solving the equation $E_0^\text{Ising} = E_0^\text{PM}$.

Now, we discuss the phase transition from the partially disordered Ising phase to the SS phase. This transition is characterized by spontaneous U(1) symmetry breaking due to the emergence of the in-plane component. The continuous transition is then determined from the softening of the  low-energy modes  of the partially disordered Ising phase: $\omega_{-,\pm \bm{Q}} = 0$. The resulting  phase boundary corresponds to the solid line in Fig.~\ref{Fig:phd}.

The phase boundary between the XY and SS phases is  denoted with a dashed line in Fig.~\ref{Fig:phd}. We note that this particular phase boundary  is calculated only at the mean-field level~\cite{Papanicolaou1988, Penc2011}.
The mean-field treatment is carried out by minimizing the energy  with respect to the variational wave function $|\Psi \rangle = \otimes_{\bm{r}} | \bm{d}_{\bm r} \rangle$, where
\begin{equation}
| \bm{d}_{\bm r}  \rangle = \frac{i d_{\bm r}^x + d_{\bm r}^y}{\sqrt{2}} | \uparrow \rangle_{\bm r}
+ \frac{-i d_{\bm r}^x + d_{\bm r}^y}{\sqrt{2}} | \downarrow \rangle_{\bm r}
-i d_{\bm r}^{z} |0\rangle_{\bm r}.
\end{equation}
Up to a U(1) rotation, the variational mean-field state for  the XY phase is:
\begin{subequations}
\begin{align}
\bm{d}_{\bm{r}_A} &= \left( 0, \, -i \sin \frac{a}{2},\, \cos \frac{a}{2}\right),\\
\bm{d}_{\bm{r}_B} &= \left( \frac{i \sqrt{3}}{2} \sin \frac{a}{2}, \, \frac{i}{2} \sin \frac{a}{2},\, \cos \frac{a}{2}\right),\\
\bm{d}_{\bm{r}_C} &= \left( \frac{-i \sqrt{3}}{2} \sin \frac{a}{2}, \, \frac{i}{2} \sin \frac{a}{2},\, \cos \frac{a}{2}\right),
\end{align}
\end{subequations}
which leads to
\begin{equation}
E_0^\text{XY} =-\frac{3J}{2} \sin^2 a + D \sin^2 \frac{a}{2}.
\end{equation}
Minimization of  $E_0^\text{XY}$ with respect to $a$ gives
\begin{equation}\label{eq:app-E0xy}
E_0^\text{XY} =  \frac{D}{2} - \frac{3J}{2} - \frac{D^2}{24J}.
\end{equation}

Up to a U(1) rotation, the variational mean-field state for the  SS phase is:
\begin{subequations}
	\begin{align}
	\bm{d}_{\bm{r}_A} &= \left( \cos \frac{a}{2}, \, i \sin \frac{a}{2} \cos b,\, i \sin \frac{a}{2} \sin b \right),\\
	\bm{d}_{\bm{r}_B} &= \left( -\cos \frac{a}{2}, \, i \sin \frac{a}{2} \cos b,\, -i \sin \frac{a}{2} \sin b \right),\\
	\bm{d}_{\bm{r}_C} &= \left( i \sin \frac{c}{2}, \, 0,\, \cos \frac{c}{2}\right),
	\end{align}
\end{subequations}
leading to
\begin{align}\label{eq:app-E0ss}
E_0^\text{SS} &= J \left( \sin a \sin b -\sin c \right)^2 - J \sin^2 c - \zeta J \sin^2 a \cos^2 b \nonumber \\
&\quad + \frac{D}{3} \left( 2 \cos^2 \frac{a}{2} + 2 \sin^2 \frac{a}{2} \cos^2 b + \sin^2 \frac{c}{2} \right) .
\end{align}
The minimum of $E_0^\text{SS}$ as a function of the three independent variatonal parameters is obtained numerically. The phase boundary between XY and SS phase results from the condition $E_0^\text{XY}  = E_0^\text{SS}$ (see the dashed line in Fig.~\ref{Fig:phd}).

\section{A complementary approach using hard-core bosons}
\label{sec:hard-core-boson}
In this appendix we discuss a complementary approach to the CL problem, which uses somewhat different transformation to hard-core bosons for $S=1$,
 but at the end leads to the same results as the approach used in the main text.

Namely, we represent spin operators at a given site via two Bose operators $a$ and $b$ ~\cite{chubukov1989,chubukov1990_1,chubukov1991,chubukov1995}
\begin{subequations}\label{ac1}
\begin{align}
S_z &= -i \left(a^\dagger b - b^\dagger a\right),  \\
S_x &= -i  \left(b^\dagger U - U b\right),  \\
S_y &= -i  \left( U a -a^\dagger U\right),
\end{align}
\end{subequations}
where $U = \left(1 - a^\dagger a - b^\dagger b\right)^{1/2}$.
The $a$ and $b$ bosons at every lattice site obey the constraint $a^\dagger a + b^\dagger b =0, 1$.
With this extra condition, spin commutation relations are satisfied, and
\begin{subequations}\label{ac2}
	\begin{align}
 S^2_z &= a^\dagger a + b^\dagger b, \\
 S^2_x &= 1- a^\dagger a, \\
 S^2_y &= 1- b^\dagger b,
	\end{align}
\end{subequations}
such that ${\bm  S}^2 = 2$, as it should be.

The Hamiltonian of Eq. (\ref{eq:ham}) is expressed via $a$ and $b$ bosons as
\beq
{\cal H} = {\cal H}_2 + {\cal H}_4,
\label{ch3}
\eeq
where in momentum space
\begin{align} \label{ch4}
{\cal H}_2  &=  \sum_{\bm k} \left[\left(D + \epsilon_{\bm k} \right) a^\dagger_{\bm k} a_{\bm k} - \frac{ \epsilon_{\bm k}}{2} \left(a^\dagger_{\bm k} a^\dagger_{-{\bm k}} + a_{\bm k} a_{-{\bm k}}\right)\right] \nonumber \\
& + \sum_{\bm k} \left[\left(D +  \epsilon_{\bm k}\right) b^\dagger_{\bm k} b_{\bm k} - \frac{ \epsilon_{\bm k}}{2} \left(b^\dagger_{\bm k} b^\dagger_{-{\bm k}} + b_{\bm k} b_{-{\bm k}}\right)\right],
\end{align}
and
\begin{align} \label{ch5}
{\cal H}_4 &=  \frac{J_z}{N} \sum_{{\bm k}_i} a^\dagger_{{\bm k}_1} b^\dagger_{{\bm k}_2} b_{{\bm k}_3} a_{{\bm k}_4} \left(\gamma_{{\bm k}_1-{\bm k}_3}  + \gamma_{{\bm k}_2-{\bm k}_4} \right)
\nn \\
&-\frac{J_z}{2N} \sum_{{\bm k}_i} \left(a^\dagger_{{\bm k}_1} a^\dagger_{{\bm k}_2}  b_{{\bm k}_3} b_{{\bm k}_4} + b^\dagger_{{\bm k}_1} b^\dagger_{{\bm k}_2} a_{{\bm k}_3} a_{{\bm k}_4} \right)  \times \nn\\
& \quad \left(\gamma_{{\bm k}_1-{\bm k}_3}  + \gamma_{{\bm k}_1-{\bm k}_4} \right).
\end{align}
Here, $\epsilon_{\bm k}= 2 J \gamma_{\bm k}$ and $\gamma_{\bm k}=\sum_{\nu}  \cos{{\bm k} \cdot {\bm e}_{\nu}}$.
There is no four-boson term from the transverse, $J$, part of the spin-spin interaction, once the constraint is satisfied.

Because the boson density  $a^\dagger a + b^\dagger b$ can have two values at a given site, there is no straightforward way to enforce the constraint by introducing the Lagrange multiplier.  One can either extend the model to $N >1$ bosonic flavors and expand in $1/N$, or just assume that the average density of bosons is small and neglect the constraint.  The last approach is rigorously justified only at large $D \gg J$, but we expect that it gives meaningful results also at $D \geq J$, as long as single-particle spin-wave excitations are gapped.  Below we just neglect the constraint and analyze the formation of the two-particle bound state within the model of Eqs. \eqref{ch4} and \eqref{ch5} with no additional constraint.   We recall in this regard that within the Schwinger boson approach, which we adopted in the main text,  we replaced one bosonic field, $b_{0}$, by its condensate value $s$ and
thereby also reduced the model to that of two interacting bosonic fields. The site-independent Lagrange multiplier $\mu$, which we introduced in the main text to enforce the constraint, and the condensate $s$ renormalize $D$ and $J$ in the quadratic form, but do not affect its structure. From this perspective, the Hamiltonian of Eqs.~\eqref{ch4} and \eqref{ch5} is  qualitatively the same as the one in the main text, assuming that one adjusts $D$ and $J$.

Furthermore, one can show that the transformation from operators $b_{\uparrow}$ and $b_{\downarrow}$ of the main text to operators
$a$ and $b$ used here is just a rotation in operator space:
\beq
a = \frac{b_{\downarrow}- b_{\uparrow}}{\sqrt{2}},~~~ b = (-i) \frac{b_{\uparrow} +b_{\downarrow}}{\sqrt{2}}
\label{ch6}
\eeq
Obviously then, the results obtained using $a$ and $b$ bosons must be equivalent to those obtained using $b_{\uparrow}$ and $b_{\downarrow}$ bosons.
 We will see, however, that technical details of the computation of the bound-state instability differ between the approaches.

We now proceed with the Hamiltonian of Eqs. (\ref{ch4}) and (\ref{ch5}).  The diagonalization of the quadratic form in \eqref{ch4} is done in the usual way. We introduce
\beq
a_{\bm k} = u_{\bm k} d_{\bm k} - v_{\bm k} d^\dagger_{-{\bm k}},~~b_{\bm k} = u_{\bm k} {\bar d}_{\bm k} - v_{\bm k} {\bar d}^\dagger_{-{\bm k}}
\label{ch6a}
\eeq
 and choose
\begin{subequations}\label{ch7}
\begin{align}
u_{\bm k} &= (D+\omega_{\bm k})/(2\sqrt{D \omega_{\bm k}}), \\
v_{\bm k} &= (D-\omega_{\bm k})/(2\sqrt{D \omega_{\bm k}}), \\
\omega_{\bm k} &= \sqrt{A^2_{\bm k} - B^2_{\bm k}} = \sqrt{D^2 + 2 D \epsilon_{\bm k}},
\end{align}
\end{subequations}
 where
 \be
A_{\bm k} = D + \epsilon_{\bm k},~~ B_{\bm k} = -\epsilon_{\bm k}.
 \label{ch8}
 \ee
The spin-wave spectrum softens at ${\bm k} = \pm {\bm Q} = \pm (4\pi/3, 0)$ at $D_c = 6J$.  In the main text, we found $D_c = 2.68 J$, which is in better agreement with the numerics. We  recall that the result was obtained by including one-loop renormalizations of $D$ and $J$. Here, we neglect these renormalizations. $D_c = 6J$ would be the critical value in the Schwinger boson analysis, presented in the main text,
if we set $s =1$ and $\mu = D$ there.

Note that \eqref{ch8} predicts that at the minimum of the magnon dispersion $\omega_{\bm Q} \sim \sqrt{D - D_c}$ while for Schwinger bosons
the relation is linear [see discussion below Eq.~\eqref{eq:scaling}].

 One can easily verify that the two-particle order parameter, which leads to a spin current state with a non-zero vector chirality $\kappa_{mn}$,
 has zero total momentum and is expressed in terms of $a$ and $b$ bosons as
  \beq
  \langle b_{\bm p} a_{-{\bm p}}\rangle = i {\tilde \Phi}_{|{\bm p}|} f_{\bm p}, ~~ \langle b^\dagger_{\bm p} a^\dagger_{-{\bm p}}\rangle = -i {\tilde \Phi}_{|{\bm p}|} f_{\bm p},
 \label{ch9}
 \eeq
  where $f_{-{\bm p}} = - f_{{\bm p}}$ is an odd function of momentum, normalized to $f_{\bm Q} =1, f_{-{\bm Q}} =-1$.
 The vector chirality on $m,n$ bond is $\kappa_{m,n} \propto \sum_{\bm p} {\tilde \Phi}_{{\bm p}}$.
 In terms of bosons $d_k$ and ${\bar d}_k$ [Eq.~\eqref{ch6a}],  the
  VC order parameter is expressed as
  \begin{subequations} \label{ch10}
  \begin{align}
 \langle {\bar d}_{\bm p} d_{-{\bm p}}\rangle &= i \Phi_{|{\bm p}|}  f_{\bm p}, \\
 \langle {\bar d}^\dagger_{\bm p} d^\dagger_{-{\bm p}}\rangle &= -i \Phi_{|{\bm p}|} f_{\bm p},
  \end{align}
  \end{subequations}
  where $\Phi_{|{\bm p}|} = {\tilde \Phi}_{|{\bm p}|}  \omega_{\bm p}/A_{\bm p}$. Note that both $\omega_{\bm p}$ and $A_{\bm p}$ are even functions of ${\bm p}$.

  \begin{figure}[tb]
  	\centering
  	\includegraphics[width=0.95\columnwidth]{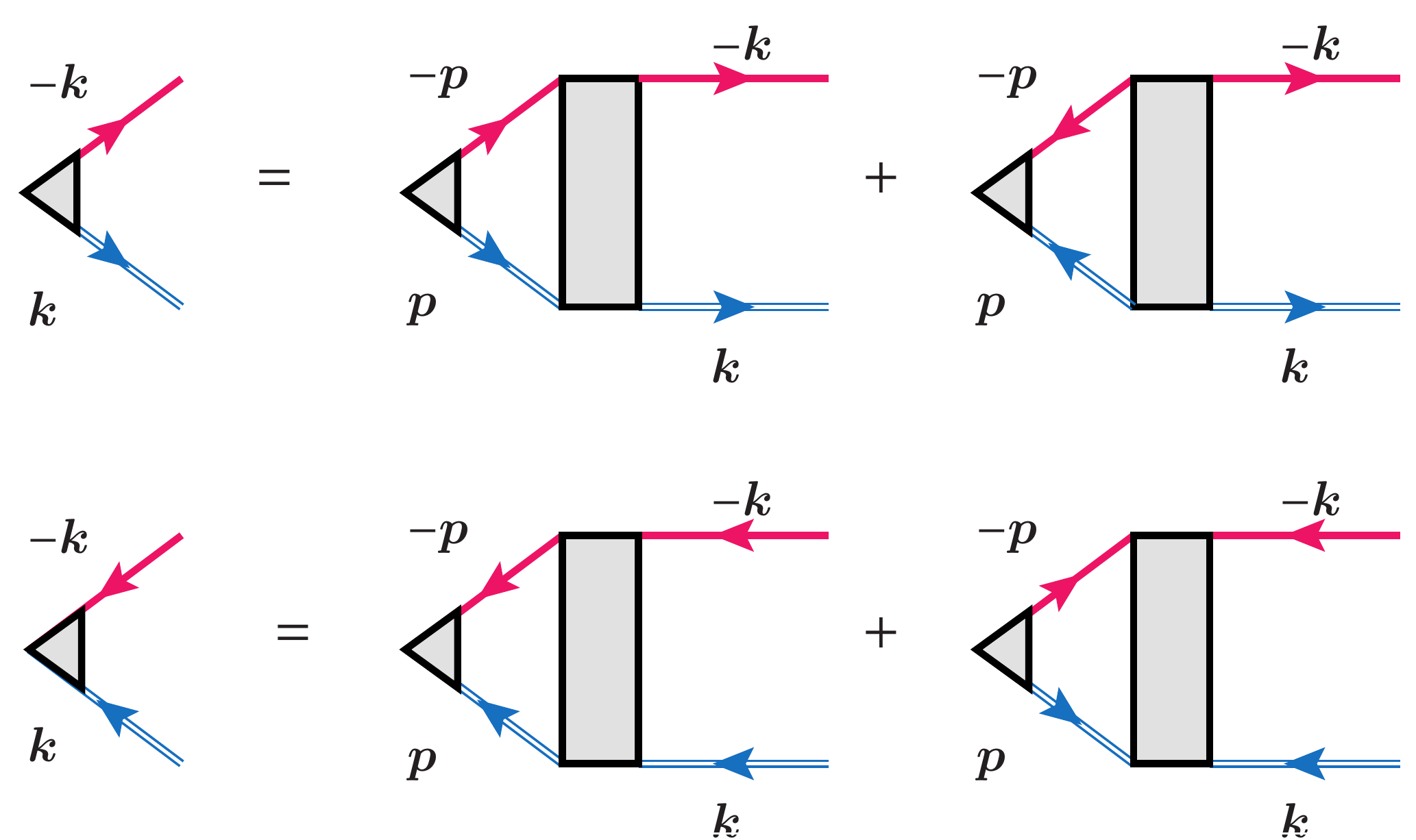}
  	\caption{Equations for
  		the  vertices $\Phi_{|{\bm p}|}$.   The triangular vertices denote $i\Phi_{|{\bm p}|}$ for ${\bm p}$ near ${\bm Q}$ and  $-i\Phi_{|{\bm p}|}$ for ${\bm p}$ near ${-\bm Q}$,   or $-i\Phi_{|{\bm p}|}$ for ${\bm p}$ near ${\bm Q}$ and  $i\Phi_{|{\bm p}|}$ for ${\bm p}$ near ${-\bm Q}$,  depending on the direction of arrows.
  		Solid  and double solid lines denote propagators of $d$ and ${\bar d}$ magnons, and the  shaded rectangles denote fully dressed irreducible interactions between low-energy magnons.}
  	\label{fig:app_1}
  \end{figure}

 We now search for the two-particle instability at $D > D_c$, i.e., preemptying to  the spin-wave instability.
 Like in the main text, we consider small $J_z = \zeta J$.
 The analysis of the two-particle instability proceeds in the same way as in Sec. \ref{sec:eom}. Namely,
  we write self-consistent equations on $\Phi_{|{\bm p}|}$ in terms of the fully renormalized irreducible pairing interaction between
  $d$ and ${\bar d}$ bosons with opposite momenta near $\pm {\bm Q}$.

  The equation for the two-particle vertex is graphically presented in Fig. \ref{fig:app_1}. It is quite similar to Fig. \ref{Fig:EOM} in the main text,
  but now the shaded triangular vertices denote  $\Phi_{|{\bm p}|}$, solid single and double lines describe propagators of $d$ and ${\bar d}$ bosons,
  and  the shaded four-point vertices represent fully dressed irreducible interactions.

   Like we said in the main text, the distinction between our problem and superconductivity is in that boson-boson interaction does not conserves the number of bosons; the interaction Hamiltonian, re-expressed in terms of bosons $d_{\bm k}$ and ${\bar d}_{\bm k}$, contains terms which create two bosons and annihilate two bosons,  and also terms which create or annihilate four bosons (as well as the terms which create three bosons and annihilate one, and vice versa).  Accordingly, the right-hand side of the equation for $\Phi$ contains both normal and ``anomalous'' terms (direction of the internal lines is the same or opposite to the direction of external lines).  At the same time, the internal part of both terms contains propagators of one $d_{\bm k}$ and one ${\bar d}_{\bm k}$ boson with the same direction of arrows.  This is because (i) bosonic dispersions are necessarily positive, hence, there is no non-zero contribution from ``particle-hole'' type terms, with different direction of arrows, and (ii) there are no graphs with two internal $d_{\bm k}$ bosons or two ${\bar d}_{\bm k}$  bosons. The latter restriction is due to the fact that $J_z$ interaction contains two $d_{\bm k}$ bosons and two ${\bar d}_{\bm k}$ bosons, simply because the original interaction \eqref{ch5} had two $a$ bosons and two $b$-bosons,
   and $a_{\bm k}$ transforms into  $d_{\bm k}$ and $b_{\bm k}$ transforms into ${\bar d}_{\bm k}$.   As a result, if external bosons are $d_{\bm k}$ and ${\bar d}_{\bm k}$ as they should be for the chiral vertex \eqref{ch10},
   one of internal bosons must be $d_{\bm k}$ and another must be  ${\bar d}_{\bm k}$.

Because the interaction vertices contain four coherence factors $u_{\bm k}$ or $v_{\bm k}$, each of which is proportional to $1/\sqrt{\omega_{\bm k}}$,
we parametrize $2 \to 2$ and $0 \to 4$ interactions between bosons with momenta $({\bm k}, -{\bm k})$ and $({\bm p},-{\bm p})$
[the analogs of $\Gamma$ terms in Eq. (\ref{Eq:EOM3})] as
\begin{subequations}\label{ch11}
\begin{align}
2 \to 2 {\rm ~interaction}:&\quad  \frac{1}{\omega_{\bm k}} \frac{1}{\omega_{\bm p}} F^{(22)} ({\bm k},{\bm p}),  \\
0 \to 4 {\rm ~interaction}:& \quad \frac{1}{\omega_{\bm k}} \frac{1}{\omega_{\bm p}} F^{(04)} ({\bm k},{\bm p}).
\end{align}
\end{subequations}
With these notations, the equation on $\Phi_{|{\bm k}|}$  takes the form
\beq
\Phi_{|{\bm k}|} \!=\! - \frac{1}{N} \sum_{\bm p} \frac{f_{\bm p}}{2 \omega^2_{\bm p} \omega_{\bm k}} \Phi_{|{\bm p}|} \left( F^{(22)} ({\bm k},{\bm p}) -  F^{(04)} ({\bm k},{\bm p})\right).
\label{ch12}
\eeq

 A technical remark: Compared to Eq. (\ref{Eq:EOM3}) in the main text, we incorporated the overall combinatoric factor of $4$ for the anomalous term into $F^{(04)}$.

 We expect, by analogy with the analysis in the main text, that $ F^{(22)} ({\bm k},{\bm p})$  and $F^{(04)} ({\bm k},{\bm p})$ are non-singular functions of momenta near ${\bm k}, {\bm p} = \pm {\bm Q}$.
 In this situation, integral equation (\ref{ch12}) can be reduced to the algebraic equation
   \beq
1 = - \frac{A}{N} \sum_{\bm p} \frac{1}{2 \omega^3_{\bm p}},
\label{ch14}
\eeq
where
\bea
A &=& \quad \left(F^{(22)} ({\bm Q},{\bm Q}) -F^{(22)} ({\bm Q},-{\bm Q})\right) \nn \\
&&-\left(F^{(04)} ({\bm Q},{\bm Q}) -F^{(04)} ({\bm Q},-{\bm Q})\right).
\label{ch14a}
\eea
We follow the analysis in the main text and consider the case when $J_z$ is small.  In this limit, both $ F^{(22)}$ and $F^{(04)}$ are obviously small in $J_z$.
  The solution of (\ref{ch14}) nevertheless seems possible because the kernel in the r.h.s. of (\ref{ch14}) contains $1/\omega^3_{\bm p}$.  Near $D = D_c$, spin-wave excitation energy $\omega_{\bm p}$  is small at ${\bm p} \approx \pm {\bm Q}$, and $\sum_{\bm p} 1/(2 \omega^3_{\bm p})$ diverges as $D$ approaches $D_c$ from above.
Then, the spin-current state emerges at arbitrary weak $J_z$ if $A$ has a finite negative value.

We now compute $A$.  To first order in $J_z$,  $F^{(22)} ({\bm k},{\bm p})$ and $F^{(04)} ({\bm k},{\bm p})$ are just the interaction terms in the Hamiltonian, re-expressed in terms of $d$ and ${\bar d}$ bosons.  Using the transformation (\ref{ch6a}) we obtain after simple algebra
\begin{subequations}\label{ch15}
\begin{align}
F^{(22)} ({\bm k},{\bm p}) &= J_z \gamma_{{\bm k}+{\bm p}} \left(A_{\bm k} A_{\bm p} - B_{\bm k} B_{\bm p} + \omega_{\bm k} \omega_{\bm p} \right),  \\
F^{(04)} ({\bm k},{\bm p}) &= J_z \gamma_{{\bm k}-{\bm p}} \left(A_{\bm k} A_{\bm p} - B_{\bm k} B_{\bm p} - \omega_{\bm k} \omega_{\bm p} \right).
\end{align}
\end{subequations}
Accordingly
\begin{subequations}\label{ch16}
\begin{align}
F^{(22)} ({\bm Q},{\bm Q}) &= 2J_z \gamma_{2{\bm Q}} \omega^2_{\bm Q},\\
F^{(22)} ({\bm Q},-{\bm Q}) &= 2J_z \gamma_{0} \omega^2_{{\bm Q}}, \\
F^{(04)} ({\bm Q},{\bm Q}) &= F^{(04)} ({\bm Q},-{\bm Q}) =0.
\end{align}
\end{subequations}

Because $\gamma_{0} =3$ and $\gamma_{2{\bm Q}} = -3/2$, the sign of $A$ is negative, i.e., the interaction in the spin-current channel is attractive.
At the same time, we see that the magnitude of $A$ scales as $ \omega^2_{\bm Q} \sim (D-D_c)$. This smallness compensates the divergence of
$(1/N) \sum_{\bm p} 1/\omega^3_{\bm p} \sim 1/\omega_{\bm Q} \sim (D-D_c)^{-1/2}$.   As a result,
Eq. (\ref{ch14}) reduces to $1= (J_z/J) (D/D_c -1)^{1/2}$, which obviously has no physical solution.
The absence of the instability in the analysis to first order in  $J_z$ agrees with the similar finding in the main text.

  \begin{figure}[tb]
	\centering
	\includegraphics[width=0.95\columnwidth]{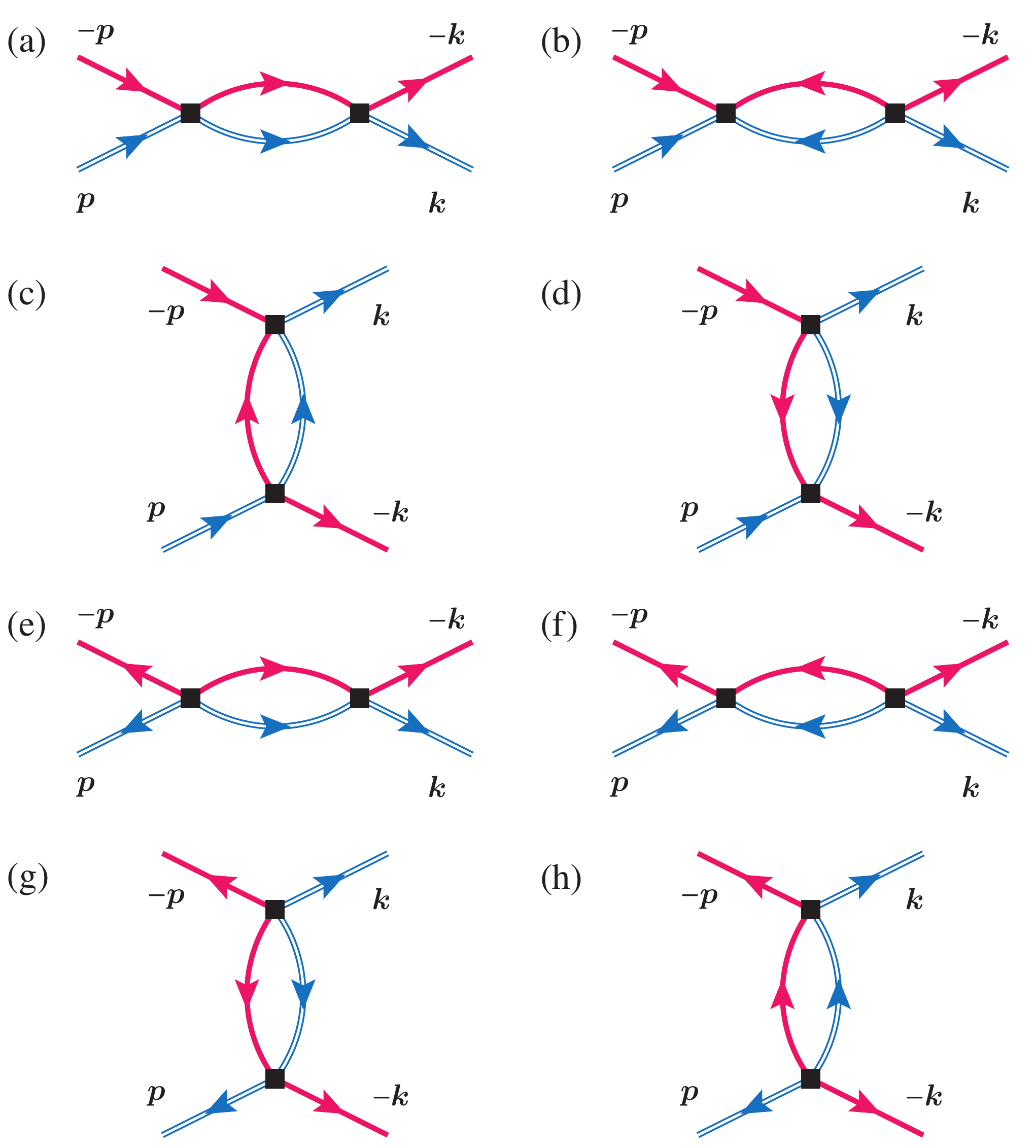}
	\caption{Equations for the dressed four-point vertices made out of $d$ and ${\bar d}$ bosons,  at the second order in $J_z$.
	(a)--(d) Normal vertices $F^{(22)}$. (e)--(h) Anomalous vertices $F^{(04)}$.}
	\label{fig:app_2}
\end{figure}

We further compute irreducible interactions to second order in $J_z$.   The corresponding contributions to $ F^{(22)} ({\bm k},{\bm p})$ and  $F^{(04)} ({\bm k},{\bm p})$ are shown in Fig.~\ref{fig:app_2}.  We set external momenta at ${\bm k} = {\bm Q}, {\bm p} = \pm {\bm Q}$, but put no restriction on internal momenta.  For practical purposes, we found it more convenient to evaluate directly the differences $\delta F^{(22)} = F^{(22)} ({\bm Q},{\bm Q}) -F^{(22)} ({\bm Q},-{\bm Q})$ and $\delta F^{(04)} = F^{(04)} ({\bm Q},{\bm Q}) -F^{(04)} ({\bm Q},-{\bm Q})$ rather than each term separately.  In these notations, $A = \delta F^{(22)} - \delta F^{(04)}$.

The contributions to irreducible $\delta F^{(22)}$ at second order in $J_z$ come from three sets of processes: the one with two $2\to 2$ interactions, the one with $0 \to 4$ and $4 \to 0$ interactions, and the one
with $1 \to 3$ and $3 \to 1$ interactions.   Each of these interaction terms is obtained from the original interaction in terms of $a$ and $b$ bosons [Eq.~\eqref{ch5}], by applying the transformation
from $a,b$ to $d, {\bar d}$ bosons [Eq.~\eqref{ch6a}]. The $2 \to 2$, $4 \to 0$, and $0 \to 4$  terms contain even numbers of $u$ and $v$ factors,
$1 \to 3$ and $3 \to 1$ terms contain either three $u$ and one $v$ factor, or vice versa.  As an example, we present the explicit expression for one of $3 \to 1$ terms:
\beq
H_{3 \to 1} = - \frac{J_z}{N} \sum_{1,2,3,4} d_{1} {\bar d}_{2} {\bar d}_{3} d_4 K_{123,4}  + ....
\label{ch17}
\eeq
 where $1 \equiv {\bm k}_1$, etc., momentum conservation is implied, and
 \bea
 K &=& \gamma_{1+3} \left[\left(u_1 v_3 - v_1 u_3\right) \left(u_2 u_4 - v_2 v_4\right)\right]   \nonumber \\
 && + \gamma_{1+2} \left[\left(u_1 v_2 - v_1 u_2\right) \left(u_3 u_4 - v_3 v_4\right)\right].
\label{ch18}
\eea
The ellipsis in (\ref{ch17}) stands for other terms with $3 \to 1$ structure.

Evaluating irreducible $\delta F^{(22)}$ and $\delta F^{(04)}$ from each of these processes and collecting combinatoric factors, we obtain
\beq
\delta F^{(22)} = - \frac{1}{4} J^2_z {\cal S}, ~~ \delta F^{(04)} =  \frac{1}{4} J^2_z  {\cal S},
\label{ch19}
\eeq
 where
\begin{align}\label{ch20}
 {\cal S} &= \frac{1}{N} \sum_{\bm k} \Bigg[\frac{\omega_{\bm k}}{2} \left(\gamma_{{\bm Q} +{\bm k}} - \gamma_{{\bm Q}-{\bm k}} \right)^2  - \omega_{\bm k} ~\gamma_{{\bm Q} +{\bm k}} \gamma_{{\bm Q}-{\bm k}}  \nonumber \\
 &\quad \qquad \quad +  \frac{2 \omega_{{\bm Q}+{\bm k}} \omega_{{\bm Q}-{\bm k}}}{ \omega_{{\bm Q}+{\bm k} } + \omega_{{\bm Q}-{\bm k}}} \gamma^2_{\bm k} \Bigg].
\end{align}
Numerical evaluation yields
${\cal S} =  12.92J$,  which is consistent with Eqs.~\eqref{Eq:normal_beta} and \eqref{Eq:anomalous_beta} in the main text: $2 {\cal S} = \sum_i \beta_i = \sum_i {\bar \beta}_i$.
Using \eqref{ch14a} and \eqref{ch19}, we obtain $A = - 6.46 J^2_z J$.
 Substituting this $A$ into Eq. \eqref{ch14}  we obtain
\begin{equation}
\frac{1}{3.23\zeta^{2} J^3}
=
\frac{1}{N}\sum_{\bm{p}}\frac{1}{\omega^{3}_{\bm p}}.
\label{nnnn_1}
\end{equation}
This is exactly the same equation as Eq. (\ref{nnnn}) in the main text, the only difference is that in the current approach  $s^2=1$.
[In Eq.~\eqref{nnnn}, the numerical factor in the left-hand side is $\alpha = 2.49 = 3.23 s^2$.]

Using the fact that $(1/N) \sum_{\bm p} 1/ \omega^{3}_{\bm p} \sim (1/J^3) (D/D_c-1)^{-1/2}$, the condition  for the instability towards the CL state
($D = D^b_c$)
takes the form
\beq
\frac{D^b_c}{D_c} - 1 \propto \left(\frac{J_z}{J}\right)^4 = \zeta^4.
\label{ch20_a}
\eeq
 Because $D^b_c > D_c$, the instability towards the CL state pre-empts the one towards the magnetically ordered XY state. This again agrees
   with the  finding in the main text.
The difference in scaling of $D_c^b - D_c$ with Ising anisotropy $\zeta$ in \eqref{eq:scaling} and \eqref{ch20_a}, $\zeta^2$ vs $\zeta^4$,
 is due to different scaling forms of the minimal magnon energy,
$\omega_{\bm Q}$, in the Schwinger boson approximation (main text) and the hard-core boson approximation  (this appendix).

\bibliography{Chiral-Loops}

\end{document}